\documentclass{ws-ijmpa}
\usepackage[super,compress]{cite}
\usepackage{graphicx}
\usepackage{enumitem}
\usepackage{hyperref}

\def\slash#1{\setbox0=\hbox{$#1$}               
        \dimen0=\wd0                            
        \setbox1=\hbox{/} \dimen1=\wd1          
        \ifdim\dimen0>\dimen1                   
        \rlap{\hbox to \dimen0{\hfil/\hfil}}    
        #1                                      
        \else              
        \rlap{\hbox to \dimen1{\hfil$#1$\hfil}} 
        /                                       
        \fi}                                    %

\begin{document}
\markboth{Daniel Pitonyak}{Transverse spin observables within collinear factorization}

%
\catchline{}{}{}{}{}
%

\title{TRANSVERSE SPIN OBSERVABLES IN HARD-SCATTERING HADRONIC PROCESSES WITHIN COLLINEAR FACTORIZATION}

\author{DANIEL PITONYAK}

\address{RIKEN BNL Research Center\\
Brookhaven National Laboratory\\
Upton, New York 11973, USA\\
dpitonyak@quark.phy.bnl.gov}

\maketitle

\begin{history}
\received{Day Month Year}
\revised{Day Month Year}
\end{history}

\begin{abstract}
We review what is currently known about the transverse spin structure of hadrons, in particular from observables that can be analyzed within a collinear framework.  These effects have been around for 40 years and represent a critical test of perturbative QCD.  We look at both proton-proton and lepton-nucleon collisions for various final states.  While the main focus is on transverse single-spin asymmetries, we also discuss how longitudinal-transverse double-spin asymmetries offer a complimentary, yet equally important, source of information on the quark-gluon content of hadrons.  We also summarize some recent progress in solidifying the theoretical formalism behind these observables and give an outlook on future directions of research.
\end{abstract}

\ccode{PACS numbers: 12.38.-t, 12.38.Bx, 13.60.-r, 13.75.Cs, 13.85.Ni, 13.85.Qk, 13.88.+e}


\section{Introduction}	

\allowdisplaybreaks

The quest to understand the spin structure of the proton (and hadrons in general) at the partonic level has been a fruitful source of research for many decades.  One of the earliest puzzles in this field was in the high-energy production of hadrons from proton-proton collisions where one of the particles involved carries a transverse polarization. In the mid-1970s, Argonne National Lab (for $p^\uparrow p\to \pi\,X$)~\cite{Klem:1976ui} and FermiLab (for $pp \to \Lambda^\uparrow X$)~\cite{Bunce:1976yb} measured large (several tens of percent) transverse single-spin asymmetries (SSAs)\footnote{FermiLab only measured the (transverse) polarization of the Lambda, but the large polarization they found would translate into a large SSA.} $A_N$, defined as
\begin{equation}
A_N \equiv \frac{d\sigma^\uparrow-d\sigma^\downarrow} {d\sigma^\uparrow+d\sigma^\downarrow} = \frac{d\sigma^L-d\sigma^R} {d\sigma^L+d\sigma^R}\,. \label{e:AN}
\end{equation} 
In Eq.~(\ref{e:AN}), $d\sigma^\uparrow\,(d\sigma^\downarrow)$ is the cross section when the hadron's (transverse) spin vector is ``up" (``down"), and $d\sigma^L\,(d\sigma^R)$ is the cross section for final-state particles going to the left (right) of the collision axis with the hadron's spin vector fixed.  Many experiments over the last 40 years confirmed the existence of large SSAs~\cite{Adams:1991rw,Krueger:1998hz,
Allgower:2002qi,Adams:2003fx,Adler:2005in,Lee:2007zzh,:2008mi,:2008qb,Adamczyk:2012qj,Adamczyk:2012xd,Bland:2013pkt,Adare:2013ekj}, and these observables are a fundamental test of perturbative QCD (pQCD).  

The initial attempt to understand SSAs within the na\"{i}ve (collinear) parton model proved unsuccessful, as one finds extremely small asymmetries in such a framework~\cite{Kane:1978nd}.  However, this calculation merely showed that SSAs are a subleading twist (twist-3) effect, i.e., one must include quark-gluon-quark ($qgq$) correlations in the description of this observable~\cite{Efremov:1981sh}.  This realization led to a rigorous formulation over the last several decades of collinear twist-3 factorization as way to analyze SSAs~\cite{Qiu:1991pp,Qiu:1991wg,Qiu:1998ia,Kanazawa:2000hz,Kouvaris:2006zy,Eguchi:2006qz,Eguchi:2006mc,Zhou:2008,Yuan:2009dw,Kang:2010zzb,Metz:2012ct,Kanazawa:2013uia,Beppu:2010qn,Koike:2011mb,Koike:2011nx,Koike:2006qv,Koike:2007rq,Koike:2009ge,Metz:2010xs,
Beppu:2013uda,Kanazawa:2014nea,Kanazawa:2015ajw,Koike:2015zya}.  In this approach, one writes the (polarized) differential cross section for $A^\uparrow+B\to C+X$ as\footnote{An analogous formula holds if C is transversely polarized instead of A.}
\begin{align} 
d\sigma(S_{T}) &= \,H\otimes f_{a/A(3)}\otimes f_{b/B(2)}\otimes D_{C/c(2)} \nonumber \\*
&+ \,H'\otimes f_{a/A(2)}\otimes f_{b/B(3)}\otimes D_{C/c(2)} \nonumber \\
&+ \,H''\otimes f_{a/A(2)}\otimes f_{b/B(2)}\otimes D_{C/c(3)}\,.
\label{e:collfac}
\end{align} 
In Eq.~(\ref{e:collfac}), $f_{a/A(t)}$ is the parton distribution function (PDF)\footnote{Although twist-3 correlators do not have a probabilistic interpretation, we will still refer to them as parton distribution functions and fragmentation functions.} associated with parton $a$ in hadron $A$ (and similarly for $f_{b/B(t)}$), while $D_{C/c(t)}$ is the fragmentation function (FF) associated with hadron $C$ in parton $c$.  The twist of the functions is indicated by $t$.  The factors $H$, $H'$, and $H''$ are the hard parts for each term, and the symbol $\otimes$ denotes convolutions in the appropriate momentum fractions. The hard scale for the reaction is set by the transverse momentum of the outgoing particle, $P_{CT}\gg \Lambda_{QCD}$, and $S_T$ is the transverse spin vector of $A$.\footnote{As we will discuss later, one also has twist-3 effects when, in addition to a transversely polarized particle, there is also a longitudinally polarized one.}  In this paper we will review how the collinear twist-3 framework outlined above has been used to describe SSAs\footnote{An alternative approach is the so-called Generalized Parton Model (GPM),~\cite{Anselmino:2012rq,Anselmino:2013rya,Anselmino:2014eza} which uses transverse momentum dependent (TMD) functions.  Such a method most likely is not valid for observables with only one large scale, and, therefore, the GPM should be considered only a phenomenological model.} (and a related double-spin asymmetry (DSA) $A_{LT}$ involving longitudinally and transversely polarized particles) in both proton-proton and lepton-nucleon collisions for various final states.  

The paper is organized as follows: in Sec.~\ref{s:colT3} we briefly review collinear twist-3 functions, which are all accessible using transverse spin observables.  In Sec.~\ref{s:ANpp} we discuss $A_N$ in $p^\uparrow p\to \{\pi,jet,\gamma\} \,X$, all of which have been measured at the Relativistic Heavy Ion Collider (RHIC), with separate subsections devoted to $\pi$ and $\{jet,\gamma\}$.  We also include a subsection on $A_{LT}$ in these reactions.  In Sec.~\ref{s:ANlN}  we talk about related observables in lepton-nucleon collisions, with emphasis their ability to help us understand proton-proton reactions and their measurability at an Electron-Ion Collider (EIC).  We dedicate Sec.~\ref{s:progress} to recent theoretical progress in solidifying the collinear twist-3 framework.  Finally, in Sec.~\ref{s:concl}  we give an outlook on future (theoretical and experimental) directions of research that will ultimately help us resolve this 40-year old puzzle of what causes SSAs.  The appendices contain the operator definitions of collinear twist-3 functions used in this paper and important relations between them.\\

\vspace{-0.5cm}
\section{Review of collinear twist-3 functions} \label{s:colT3}
In this section we give an overview of collinear twist-3 functions, specifically those in the quark sector, i.e., quark-quark and quark-gluon-quark correlators.\footnote{The gluon sector of twist-3 functions is also relevant for the observables we will analyze. 
However, these correlators do not give significant contributions in the forward (large $x_F$) region\cite{Beppu:2013uda} where the asymmetries we will discuss, like $A_N$ in $p^\uparrow p \to\pi\,X$, are greatest.} We focus on conveying the landscape of available functions and leave the rigorous operator definitions for \ref{a:def}.  
\begin{table}[h]
\tbl{Collinear twist-3 functions for unploarized (U), longitudinally polarized (L), and transversely polarized (T) hadrons.}
{\centering \includegraphics[scale=0.45]{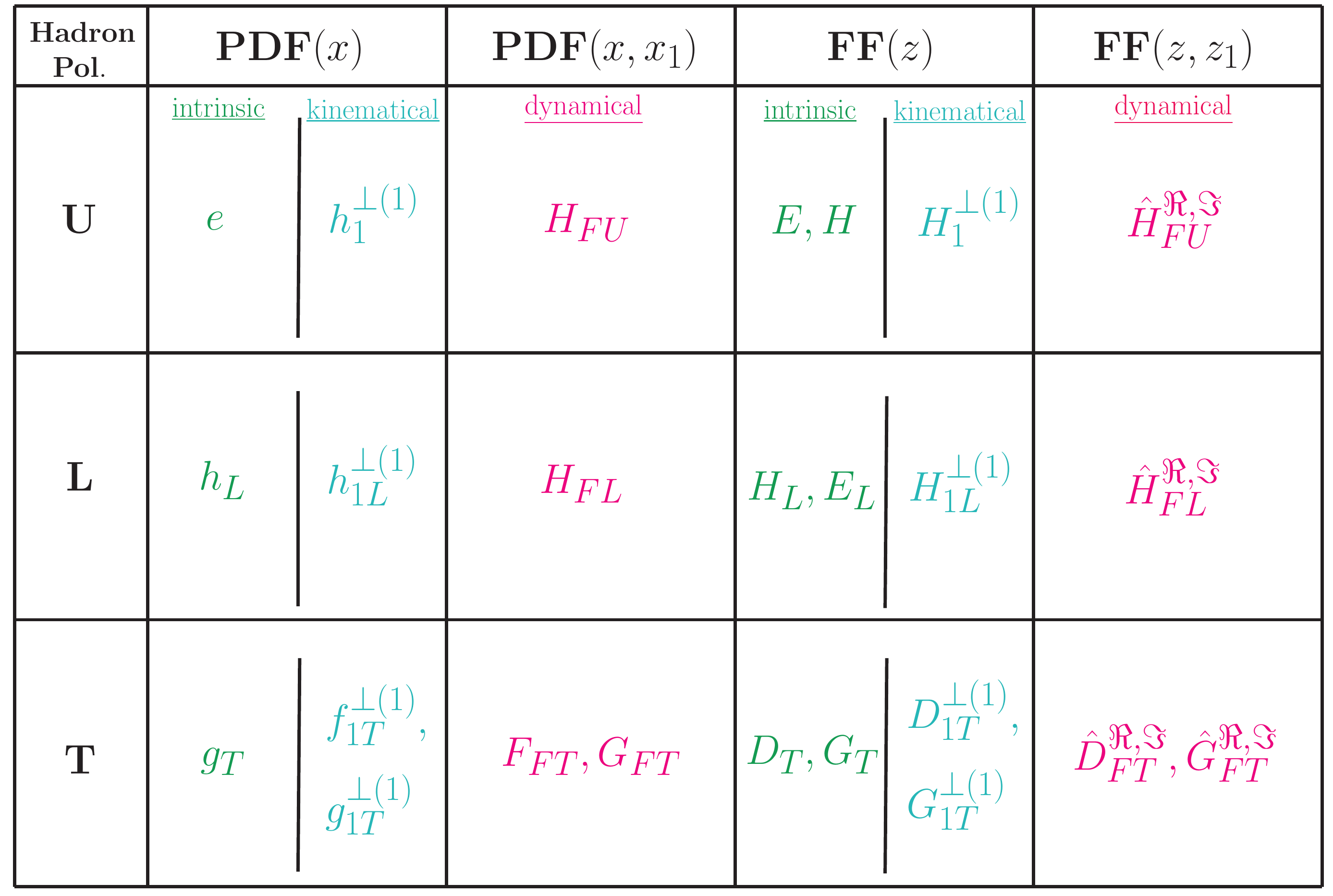}} \label{t:T3func}
\vspace{0.3cm}
\end{table} 
As we allude to at several points throughout the paper, and elaborate on more in Sec.~\ref{s:progress}, all of the collinear twist-3 functions presented here are not independent of each other.  
There are formulae, called QCD equation of motion (EOM) relations and Lorentz invariance relations (LIRs), that connect specific sets of functions.  
We write these out explicitly in \ref{a:EOM} and \ref{a:LIR}.  
We mention that various names, notations, and definitions for collinear twist-3 functions are used in the literature, and we will follow those of Ref.~\refcite{Kanazawa:2015ajw}. Note that certain subsets of these correlators are also discussed in Refs.~\citen{Qiu:1998ia,Kouvaris:2006zy,Eguchi:2006qz,Eguchi:2006mc,Koike:2009ge,Beppu:2010qn,Koike:2011mb,Koike:2011nx,Beppu:2013uda,Kanazawa:2000hz,Kanazawa:2014nea,Koike:2015zya,Kang:2010zzb,Metz:2012ct,Kanazawa:2013uia}. 
\begin{figure}[t]
\centering \includegraphics[scale=0.5]{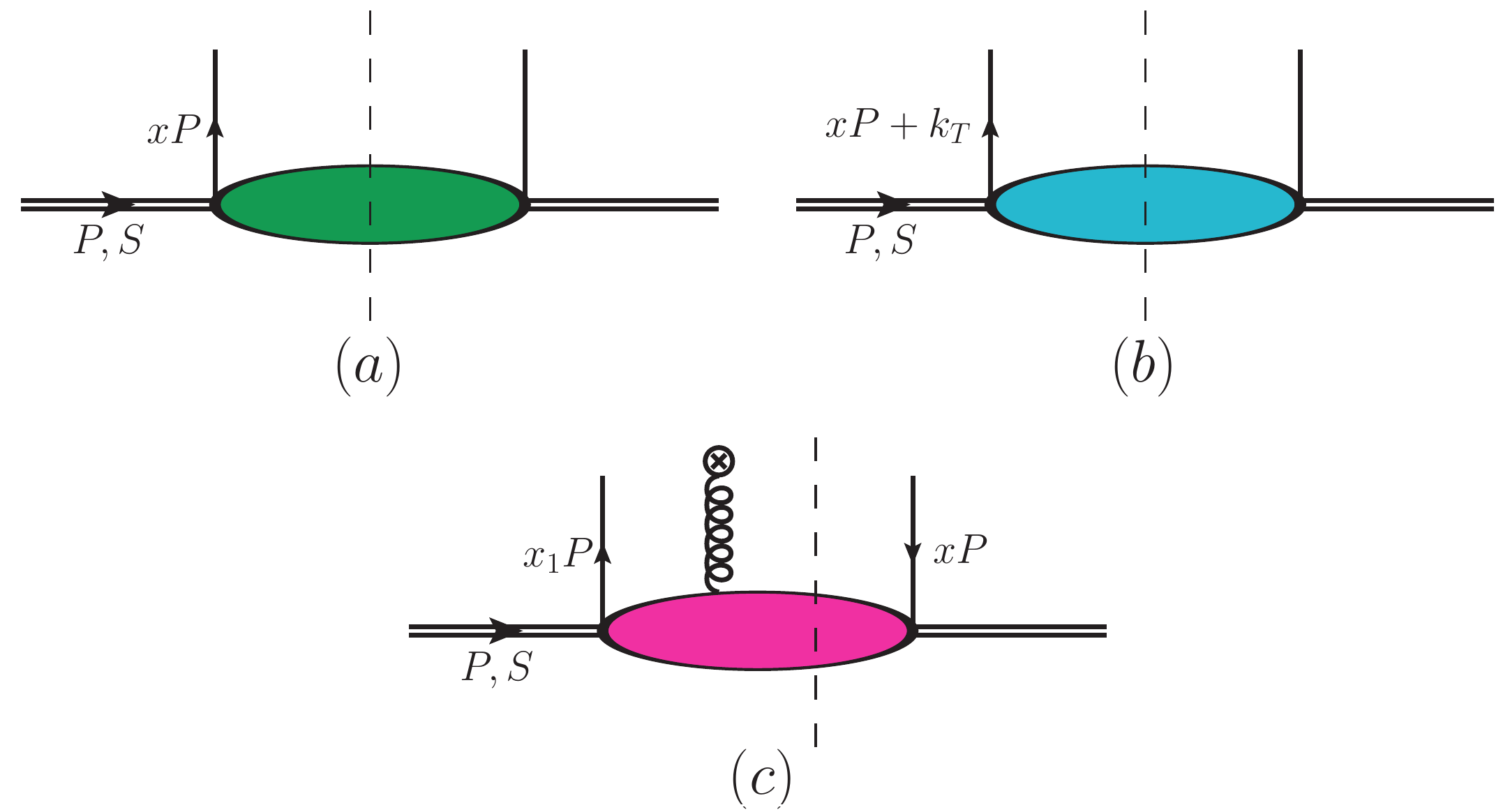}
\vspace{-0.3cm}
\caption{Feynman diagrams for (a) intrinsic, (b) kinematical, and (c) dynamical twist-3 PDFs. } \label{f:T3matrix}
\vspace{-0.3cm}
\end{figure}

\begin{figure}[t]
\centering \includegraphics[scale=0.5]{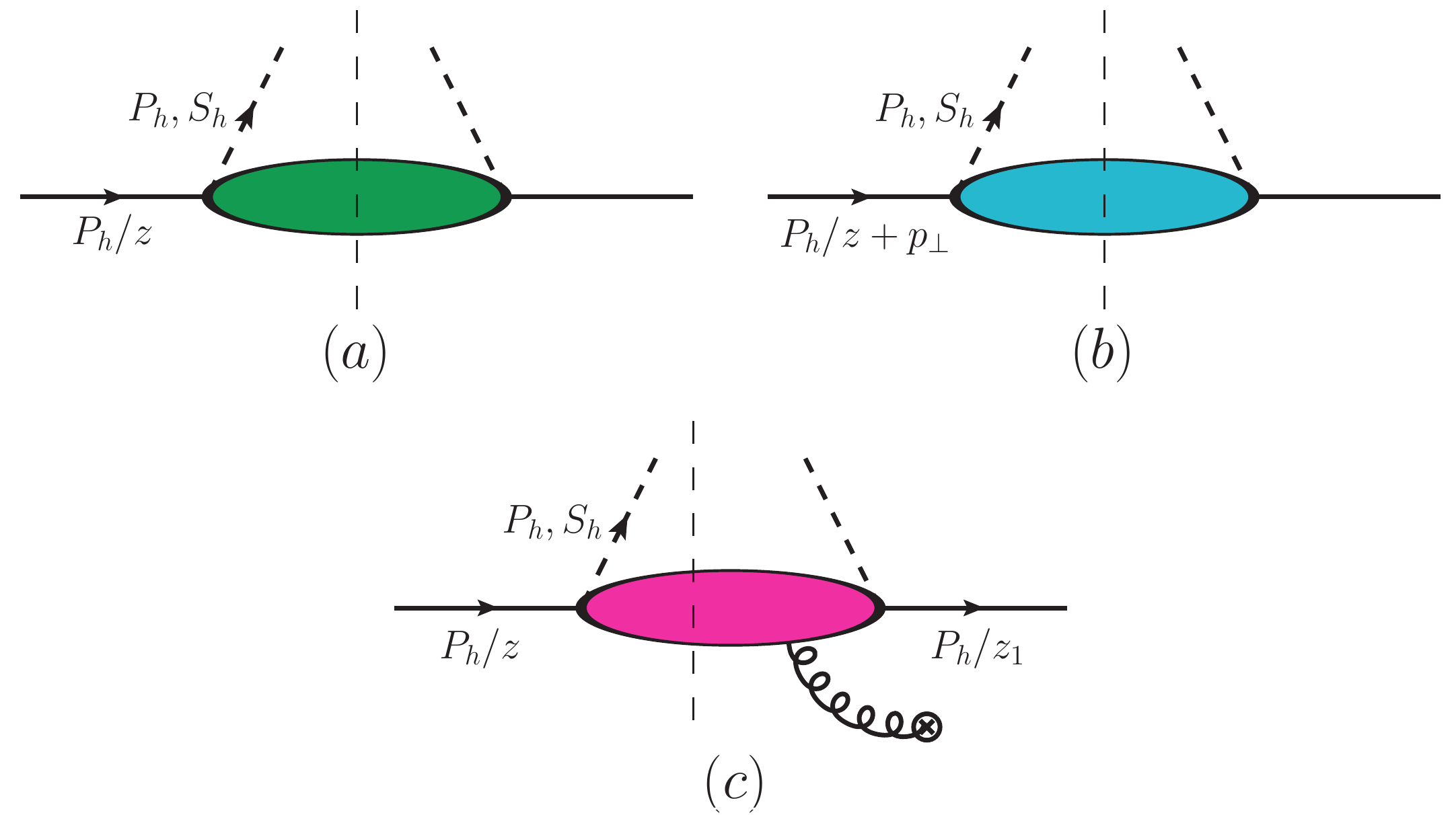} \vspace{-0.3cm}
\caption{Feynman diagrams for (a) intrinsic, (b) kinematical, and (c) dynamical twist-3 FFs. } \label{f:T3matrixFrag}
\end{figure}

The full list of twist-3 functions (in the quark sector) is given in Table~\ref{t:T3func} for both PDFs and FFs, categorized in terms of hadron polarization (U = unpolarized, L = longitudinal, T = transverse).  
As seen in the table, we can organize the functions into three groups:~intrinsic, kinematical, and dynamical.~\cite{Kanazawa:2015ajw}  First, the {\it intrinsic} functions are twist-3 Dirac projections of collinear quark-quark correlators (see Figs.~\ref{f:T3matrix}(a), \ref{f:T3matrixFrag}(a)).  Next, the {\it kinematical} functions are first ($k_T$ or $p_\perp$) moments of transverse momentum dependent (TMD) functions (see Figs.~\ref{f:T3matrix}(b), \ref{f:T3matrixFrag}(b)), defined as
\begin{align}
f^{(1)}(x) &= \int\!d^2k_T\frac{\vec{k}_T^{\,2}} {2M^2}\,f(x,k_T^2)\,,\label{e:kTmom}\\
D^{(1)}(z) &=z^2\int\!d^2p_\perp\frac{\vec{p}_\perp^{\,2}} {2M_h^2}\,D(z,z^2p_\perp^2)\,. \label{e:pTmom}
\end{align}
Finally, the {\it dynamical} functions are quark-gluon-quark correlators (see Figs.~\ref{f:T3matrix}(c), \ref{f:T3matrixFrag}(c)), where the gluon field can either be written in terms of the field strength tensor $F^{\mu\nu}$ (so-called {\it F-type} functions) or the covariant derivative $D^\mu$ (so-called {\it D-type} functions).  We only list the F-type functions since the D-type can be written in terms of them (see \ref{a:FD}).  Note that both the intrinsic and kinematical twist-3 functions depend on a single momentum fraction ($x$ or $z$), whereas the dynamical correlators depend of two ($x,x_1$ or $z,z_1$).  We also mention that, while the dynamical PDFs are purely real, the dynamical FFs are complex (due to the lack of a time-reversal constraint), and we indicate their real and imaginary parts by $\Re$ and $\Im$ superscripts, respectively.  The wealth of functions shown in Table~\ref{t:T3func} demonstrates the rich structure inside of hadrons, which, as we will see, can be probed through various transverse spin observables.

\section{$A_N$ in proton-proton collisions} \label{s:ANpp}

\subsection{$p^\uparrow \!\!\!p\to \pi\,X$}
We consider SSAs in the single-inclusive production of pions from proton-proton collision, 
\begin{equation}
p(P,S_P) + p(P') \rightarrow \pi(P_h) + X\,, 
\end{equation}
where we have indicated the momenta and polarizations of the particles.  We also define the Mandelstam variables $S,T,U$ as
\begin{equation} \label{e:Mand}
S = (P+P')^2\,, \quad T = (P-P_h)^2\,, \quad U = (P'-P_h)^2\,.
\end{equation}
In this case, all three terms in Eq.~(\ref{e:collfac}) enter into the analysis.  Specifically, one receives twist-3 contributions from (a) the transversely polarized proton, (b) the unpolarized proton, and (c) the (unpolarized) final-state pion.  As we discussed in Sec.~\ref{s:colT3}, all of these pieces involve 2-parton and 3-parton correlation functions.

For (a), there are two types of terms that arise, a so-called soft-gluon pole (SGP) term and a soft-fermion pole (SFP) term.  These are so named because, since SSAs are a na\"{i}ve time-reversal odd (T-odd) effect, one must pick up a pole in the hard scattering. This pole causes the momentum fraction of either a gluon or quark in the multi-parton correlator to vanish, which leads, respectively, to a SGP or SFP.  The SGP term was calculated in Refs.~\citen{Qiu:1998ia,Kouvaris:2006zy} for $qgq$ correlators and Ref.~\refcite{Beppu:2013uda} for tri-gluon ($ggg$) ones, while the SFP term was computed in Ref.~\refcite{Koike:2009ge}.  

For many years it was thought that the $qgq$ SGP function $F_{FT}(x,x)$, called the Qiu-Sterman (QS) function,\footnote{There are several notations use in the literature for the QS function, e.g., $T_F(x,x)$ and $G_F(x,x)$.} was the dominant source of $A_N$ in this reaction~\cite{Qiu:1998ia,Kouvaris:2006zy}.  The contribution from this correlator to the spin-dependent cross section reads~\cite{Qiu:1998ia,Kouvaris:2006zy}
\begin{align}
\label{finalcr}
\frac{E_hd\sigma_{(T)}^{SGP_{qgq}}(S_P)} {d^{3}\vec{P}_{h}}
&= -\frac{4\alpha_S^2 M} {S}\,\epsilon^{P'\!PP_h S_P}\,\sum_i\sum_{a,b,c}\int_0^1\!\frac{dz} {z^3}\int_0^1 \!dx'\int_0^1 \!dx\,\delta(\hat{s}+\hat{t}+\hat{u})\nonumber\\
&\times\, \frac{\pi} {\hat{s}\hat{u}} \,f_1^b(x')\,D_1^c(z)\left[F_{FT}^a(x,x)-x\frac{dF_{FT}^a(x,x)} {dx}\right]S^i_{F_{FT}}\,,
\end{align}
where $\sum_i$ is a sum over all partonic interaction channels, $M$ is the proton mass, $\alpha_s = g^2/4\pi$ with $g$ the strong coupling, $f_1$ ($D_1$) is the standard twist-2 unpolarized PDF (FF), and the Levi-Civita tensor is defined with $\epsilon^{0123} = +1$.  The hard factors are denoted by $S_{F_{FT}}^i$ and can be found in Ref.~\refcite{Kouvaris:2006zy}.  They are functions of the partonic Mandelstam variables
\begin{equation} \label{e:pMand}
\hat{s} = xx' S\,, \quad \hat{t} = xT/z\,, \quad \hat{u} = x'U/z\,.
\end{equation}
The notation used for the cross section indicates that this is the $qgq$ SGP term for the transversely polarized proton.

The QS function has an important, model-independent relation to the TMD Sivers function~\cite{Sivers:1989cc} $f_{1T}^{\perp}(x,k_T^2)$ that enters SSAs in processes like semi-inclusive deep-inelastic scattering (SIDIS) and Drell-Yan (DY).  The identity reads~\cite{Boer:2003cm}
\begin{equation}
 \pi F^q_{FT}(x,x) =  f_{1T}^{\perp(1),q}(x)\big|_{SIDIS} = -f_{1T}^{\perp(1),q}(x)\big|_{DY}\,.
 \label{e:QS_Siv}
\end{equation}
where
\begin{equation}
f_{1T}^{\perp(1),q}(x)\equiv\!\int\!d^2k_T \,\frac{\vec{k}_T^2} {2M^2} f_{1T}^\perp(x,k_T^2)\,.
\label{e:first_mom}
\end{equation}
The second equality in Eq.~(\ref{e:QS_Siv}) reflects the well-known non-universality of the Sivers function, i.e., the function changes sign from SIDIS to DY due to the fact that the TMD correlators for those processes differ in their gauge-link structure~\cite{Collins:2002kn}.  This process dependence is a {\it prediction} from our current understanding of QCD and TMD factorization and still must be verified experimentally.  Nevertheless, there is already evidence from theory~\cite{Metz:2012ui,Gamberg:2013kla} and a recent STAR measurement~\cite{Adamczyk:2015gyk} that this sign change is correct.

From Eq.~(\ref{e:QS_Siv}), one can see that there are two ways to access the QS function: extract $F_{FT}(x,x)$ directly from data on $A_N$ in $p^\uparrow p\to \pi\,X$~\cite{Kouvaris:2006zy,Kanazawa:2010au} (again, assuming only the $qgq$ SGP term in (a) causes the asymmetry)\footnote{Note that Ref.~\refcite{Kanazawa:2010au} also included the SFP term.} or take an extraction of $f_{1T}^{\perp}(x,k_T^2)$ from SIDIS data on the Sivers asymmetry~\cite{Anselmino:2008sga} and calculate the r.h.s.~of Eq.~(\ref{e:first_mom}).  The former should agree with the latter.  However, the authors of Ref.~\refcite{Kang:2011hk} realized that these two approaches do {\it not} yield the same result:~the two different extractions disagree in sign.  This discrepancy became known as the ``sign mismatch'' crisis~\cite{Kang:2011hk}.  There were attempts to resolve this issue by allowing for more flexible parameterizations of the Sivers function, but these proved unsuccessful.~\cite{Kang:2012xf} In fact, the authors of Ref.~\refcite{Metz:2012ui} have shown compelling evidence from SSAs in inclusive DIS~\cite{Airapetian:2009ab,Katich:2013atq} that the QS function {\it cannot} be the main cause of $A_N$ in $p^\uparrow p\to \pi\,X$.  Also, the tri-gluon term has been shown to give small effects in the forward region~\cite{Beppu:2013uda}, while the SFP piece might play some role, although it will not be able to account for all of the asymmetry~\cite{Kanazawa:2010au,Kanazawa:2011bg}.

Thus, one must calculate (b) and (c) to see if either of these pieces can cause the pion SSAs.  The case of twist-3 effects in the unpolarized proton was analyzed many years ago in Ref.~\refcite{Kanazawa:2000hz}, and they were found to be negligible.   More recently, twist-3 effects due to the final-state pion were computed in Ref.~\refcite{Metz:2012ct}, which marked the first complete calculation of this term,\footnote{The so-called derivative term was already calculated in Ref.~\refcite{Kang:2010zzb}.} and the result reads~\cite{Metz:2012ct}
\begin{align}
\frac{E_hd\sigma^{Frag}(S_P)} {d^{3}\vec{P}_{h}} &= -\frac{4\alpha_{s}^{2}M_{h}} {S}\, \epsilon^{P'\!PP_h S_P}\sum_{i}\sum_{a,b,c}\int_{0}^{1}\frac{dz} {z^{3}} \int_{0}^{1}\!dx' \int_0^1 \!dx\,\delta(\hat{s}+\hat{t}+\hat{u})\,\nonumber\\ 
&\hspace{-2.2cm}\times\,\frac{1} {\hat{s}\,(-x'\hat{t}-x\hat{u})} \,h_{1}^{a}(x)\,f_{1}^{b}(x')\left\{\left[H_1^{\perp(1),c}(z)-z\frac{dH_1^{\perp(1),c}(z)} {dz}\right]S_{H_1^{\perp}}^{i} + \frac{1} {z} H^{c}(z)\, S_{H}^{i}\right.\nonumber \\[-0.1cm]
& \hspace{3cm}+\, \frac{2} {z}\int_z^\infty\! \frac{dz_1} {z_1^2}\frac{1} {\left(\frac{1} {z}-\frac{1} {z_{1}}\right)^{\!2}}\, \hat{H}_{FU}^{c,\Im}(z,z_{1}) \,S_{\hat{H}_{FU}}^{i}\Bigg\},
\label{e:sigmaFrag}
\end{align}
where $M_h$ is the pion mass, $h_1$ is the standard twist-2 transversity function, and the hard factors for each term, which can be found in Ref.~\refcite{Metz:2012ct}, are represented by $S^i$.  The notation for the cross section indicates that this is the entire (unpolarized) fragmentation term.  The functions $H_1^{\perp(1)}, H, \hat{H}_{FU}^\Im$ are unpolarized twist-3 FFs, with the first two given by quark-quark correlators and the last one by (the imaginary part of) a $qgq$ matrix element (cf.~Table \ref{t:T3func}).\footnote{Note that, unlike the QS term, the $qgq$ FF $\hat{H}_{FU}^\Im$ involves a non-pole matrix element. (The pole pieces actually vanish~\cite{Meissner:2008yf,Kanazawa:2015ajw}.)}  These twist-3 FFs are related to each other through the QCD equation of motion (EOM) relation (cf.~Eq.~(\ref{e:EOM1FF})),~\cite{Kang:2010zzb,Metz:2012ct,Kanazawa:2013uia}
\begin{equation}
H^{q}(z) \! = \! - 2 z H_1^{\perp(1),q}(z) 
           + 2 z \! \int_{z}^{\infty} \frac{dz_1} {z_1^2} \!\! \frac{1} {\frac{1} {z}-\frac{1} {z_{1}}} \hat{H}_{FU}^{q,\Im}(z,z_{1}) \,.
\label{e:EOM_FF}
\end{equation}
The first term on the r.h.s.~of (\ref{e:EOM_FF}) is the first $p_\perp$-moment of the TMD Collins FF $H_{1}^{\perp}(z,z^2p_{\perp}^{2})$ that enters SSAs in SIDIS and electron-positron annihilation $e^+e^-\!\to h_1\,h_2 \,X$:
\begin{equation}
H_1^{\perp(1),q}(z) \equiv z^2\int \!d^2 p_{\perp} \, \frac{\vec{p}_{\perp}^{\,2}}{2 M_h^2} \, 
H_{1}^{\perp,q}(z,z^2p_{\perp}^{2})\,.
 \label{e:H1perp}
\end{equation}
Therefore, one can use an extraction of the Collins function from SIDIS and $e^+e^-$ data~\cite{Anselmino:2013vqa} to fix $H_1^{\perp(1),q}(z) $.

\begin{figure}[t]
\centering \includegraphics[scale=0.80]{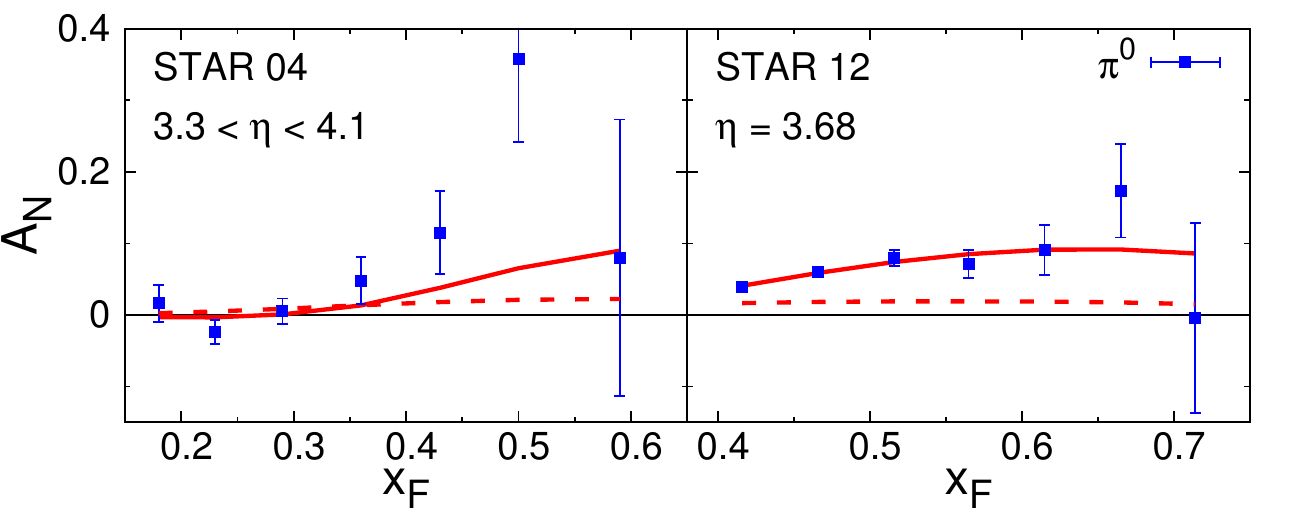} \\
\includegraphics[scale=0.80]{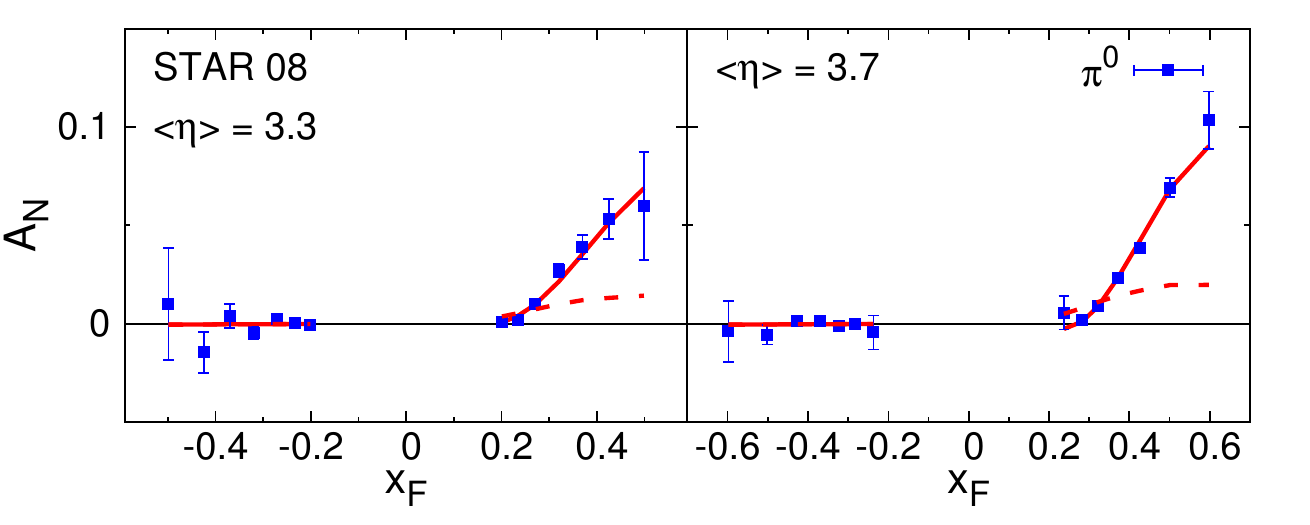} \\
\hspace{0.1cm}\includegraphics[scale=0.80]{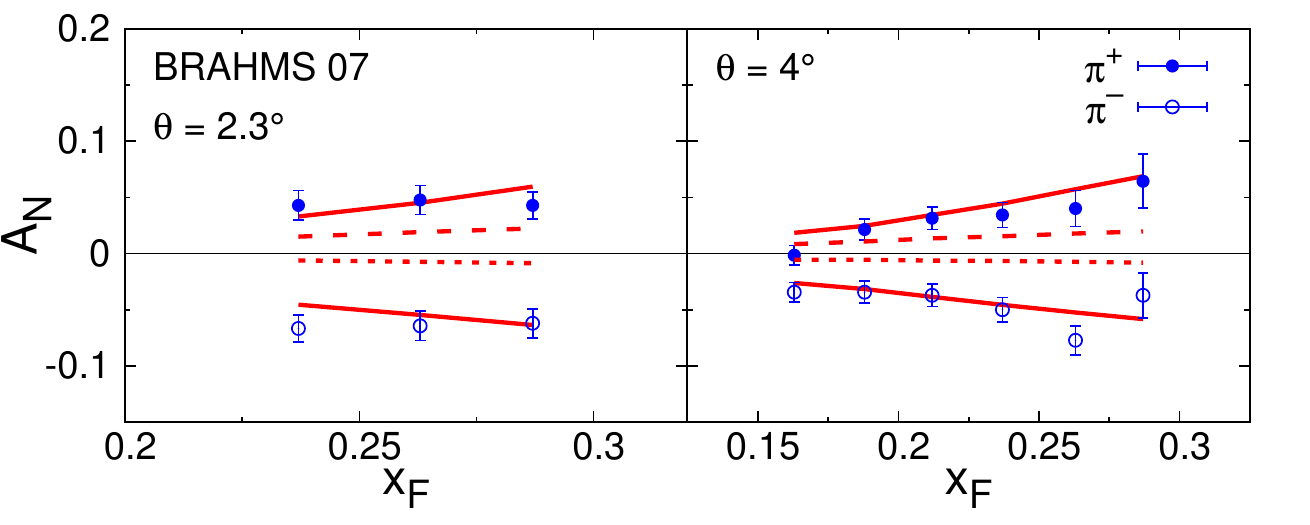}
\vspace{-0.3cm}
\caption{Fit results for $A_N$ in  $p^\uparrow p\to \pi^0\,X$ (data from Refs.~\cite{Adams:2003fx,:2008qb,Adamczyk:2012xd}) and $A_N$ in  $p^\uparrow p\to \pi^\pm\,X$ (data from Ref.~\cite{Lee:2007zzh}) for the SV1 input. The dashed line (dotted line in the case of $\pi^-$) means $\hat{H}_{FU}^{\Im}$ switched off.  Figure reprinted with permission from \href{http://dx.doi.org/10.1103/PhysRevD.89.111501} {K.~Kanazawa, et al., Phys.~Rev.~D89, 111501(R) (2014)}.  Copyright (2014) by the American Physical Society.} \label{f:fit}
\vspace{-0.3cm}
\end{figure}

\begin{figure}[t]
\centering
\includegraphics[scale=0.80]{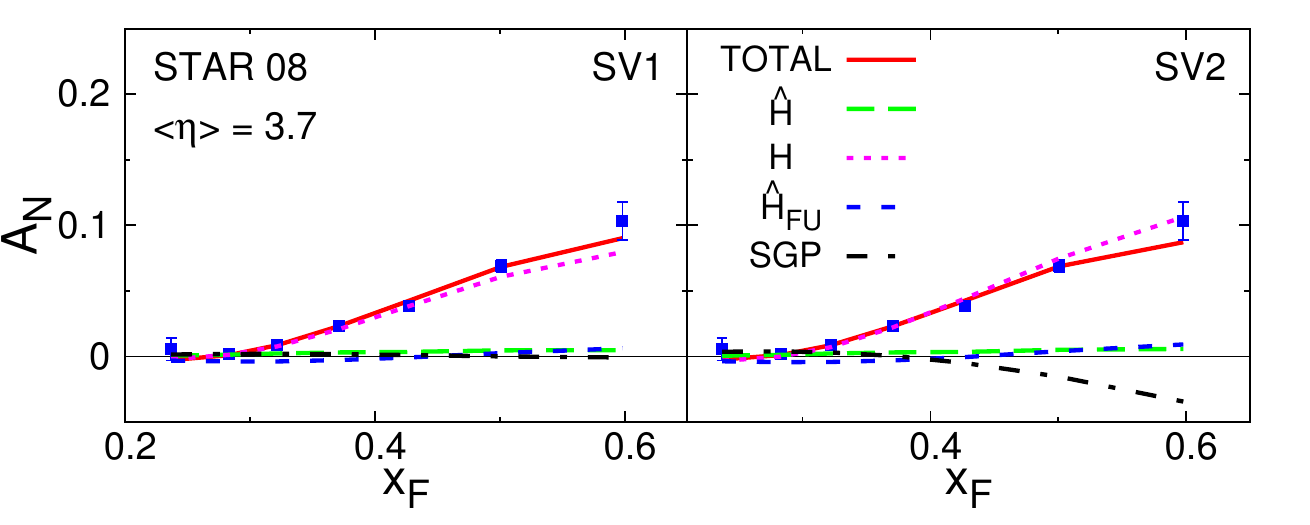}
\vspace{-0.3cm}
\caption{Individual contributions to $A_N$ in $p^\uparrow p\to \pi^0\,X$ (data from Ref.~\cite{:2008qb}) for SV1 and SV2 inputs.  Figure reprinted with permission from \href{http://dx.doi.org/10.1103/PhysRevD.89.111501} {K.~Kanazawa, et al., Phys.~Rev.~D89, 111501(R) (2014)}.  Copyright (2014) by the American Physical Society.}
\label{f:contrib}
\vspace{-0.3cm}
\end{figure}

\begin{figure}[t]
\centering
\includegraphics[scale=0.8]{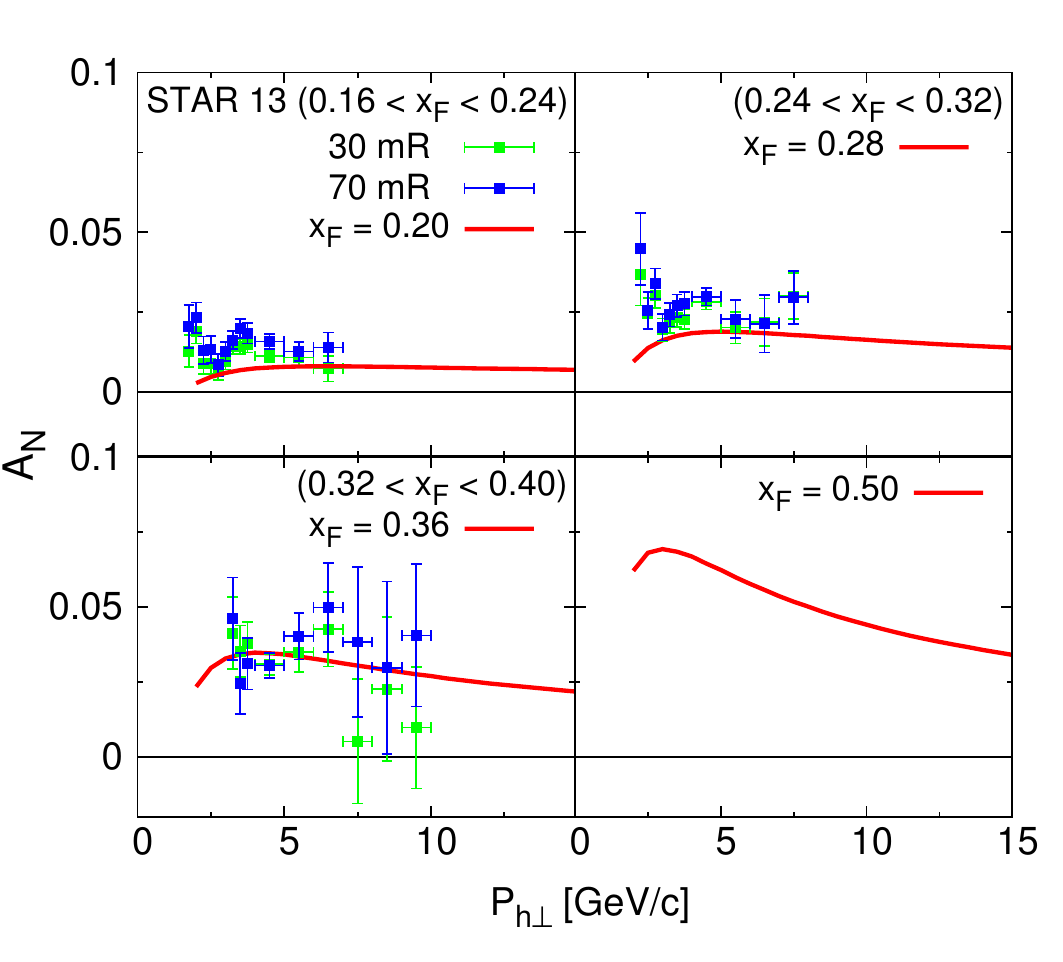}
\vspace{-0.3cm}
\caption{$A_N$ vs.~$P_{hT}$ in $p^\uparrow p\to \pi^0\,X$ for SV1 input at $\sqrt{S} = 500 \, \textrm{GeV}$ (data from Ref.~\cite{Heppelmann:2013ewa}).  Figure reprinted with permission from \href{http://dx.doi.org/10.1103/PhysRevD.89.111501} {K.~Kanazawa, et al., Phys.~Rev.~D89, 111501(R) (2014)}.  Copyright (2014) by the American Physical Society.}
\label{f:pt}
\vspace{-0.3cm}
\end{figure}

Still, in order to obtain a numerical result for this term, an input for the FF $\hat{H}_{FU}^\Im$ is required, which then allows $H$ to be determined through Eq.~(\ref{e:EOM_FF}).\footnote{For a model calculation of these functions, see Ref.~\refcite{Lu:2015wja}.}  The authors of Ref.~\refcite{Kanazawa:2014dca} parameterized $\hat{H}_{FU}^\Im$ in terms of the standard twist-2 unpolarized FF $D_1$ and also included the SGP term from Eq.~(\ref{finalcr}) with $F_{FT}(x,x)$ fixed by the Sivers function through Eq.~(\ref{e:QS_Siv}).\footnote{Two different parameterizations of the Sivers function were used, denoted SV1~\cite{Anselmino:2008sga} and SV2~\cite{Anselmino:2013rya} .}  
The parameters were fit to the RHIC $A_N$ data for charged and neutral pions~\cite{Adams:2003fx,:2008qb,Adamczyk:2012xd,Lee:2007zzh}.  The results are shown in Figs.~\ref{f:fit}--\ref{f:pt}.\footnote{Note that Ref.~\refcite{Kanazawa:2014dca} uses the notation $\hat{H}(z)$ for $H_1^{\perp(1)}(z)$.}  We see from Fig.~\ref{f:fit} that there is a very good description of the data.  In particular, from Fig.~\ref{f:contrib} one can conclude that the cause of the asymmetry is from parton fragmentation effects during the formation of the final-state pion, with the main source coming from the $qgq$ FF $\hat{H}_{FU}^\Im$.  Note that the SGP term (from the QS function) is opposite in sign to the data, and the fragmentation piece must ``overcome'' that contribution.  In addition, we see from Fig.~\ref{f:pt} that the phenomenological results agree with the flat $P_{hT}$ dependence measured at RHIC.  This was the first time that $A_N$ in $p^\uparrow p\to \pi\,X$, calculated within the collinear factorization framework, matched the RHIC data without any sign-mismatch problem, and, moreover, was consistent with TMD observables in SIDIS and $e^+e^-$~\cite{Kanazawa:2014dca}.  Nevertheless, there is still more work required in order to definitively conclude that these SSAs are due to fragmentation effects.


\subsection{$p^\uparrow \!\!\!p\to \{jet,\gamma\}\,X$}
As is clear from the last subsection, there are several pieces that enter into SSAs for pion production.  In order to better understand the mechanism that causes these asymmetries, it would be helpful to isolate certain terms and study them individually.  For the case of $A_N$ in $p^\uparrow p\to jet\,X$ and $p^\uparrow p\to \gamma\,X$, one can eliminate contributions from twist-3 FFs (i.e., the third term in Eq.~(\ref{e:collfac})) and focus on effects from the incoming protons.  However, even in this situation, one can only gain information on a specific non-perturbative function if the term involving that function is dominant.  In particular, it would be beneficial to isolate the QS function in order to see if the subsequent extraction satisfies the identity (\ref{e:QS_Siv}).  This would require a reaction where the (unpolarized and transversely polarized) SFPs, unpolarized $qgq$ SGPs, and tri-gluon SGPs are negligible (in addition to there being no fragmentation effects).  Note that one is able to derive Eq.~(\ref{e:QS_Siv}) because of the gauge link in the Sivers function, which directly causes its predicted process dependence.  Therefore, a ``clean'' extraction of the QS function can give us strong evidence of whether this non-universality is correct.

We will first consider if the jet SSA meets the above criteria to isolate $F_{FT}(x,x)$.  This asymmetry was measured by the A$_N$DY Collaboration and found to be very small ($< 0.5\%$) except for the most forward and backward $x_F$ values, where the asymmetry could be on the order of $\sim 1\%$.~\cite{Bland:2013pkt}  The analytical result for $A_N$ in jet production is obtained from the pion case by simply setting $D_1(z) = \delta(1-z)$ in the first and second terms of Eq.~(\ref{e:collfac}).  This asymmetry has been investigated in Refs.~\refcite{Kouvaris:2006zy,Gamberg:2013kla,Kang:2011hk,Kanazawa:2012kt}. Unfortunately, the conclusion as to which piece dominates is not obvious.  The study in Ref.~\refcite{Kanazawa:2000hz} provides evidence (from the pion $A_N$) that unpolarized $qgq$ SGPs and SFPs should be small in the whole $x_F$-region.  The work in Ref.~\refcite{Kanazawa:2012kt} shows the same is most likely true for transversely polarized SFPs, but that analysis suffers from the sign-mismatch issue.  Also, in Ref.~\refcite{Beppu:2013uda} there is an indication that tri-gluon SGPs could be significant.  Therefore, it will be necessary to re-assess the impact of these pieces on the jet SSA.  Nevertheless, one can still gain insight into these other terms by looking at the contribution solely from the QS function and comparing it with data.  This was done in Ref.~\refcite{Gamberg:2013kla}, where $F_{FT}(x,x)$ was fixed through the Sivers function and the result was compared to the A$_N$DY data on the asymmetry.  Although the phenomenology is in reasonable agreement with the data, there is still some room for other pieces to contribute.  Thus, $A_N$ in jet production may not be the cleanest way to access the QS function.

There is also data on tape for the prompt photon asymmetry,\footnote{There are two types of prompt photons: those that are emitted directly from the hard scattering (called direct photons) and those that form during parton fragmentation (called fragmentation photons).  We focus on direct photon production and refer the reader to Ref.~\refcite{Gamberg:2012iq} for an analysis that includes fragmentation photons.  We note that from an experimental standpoint, fragmentation photons can largely be eliminated through isolation cuts.} 
\begin{equation}
p(P,S_P) + p(P') \rightarrow \gamma(q) + X\,, 
\end{equation}
from both the PHENIX Collaboration \cite{PHENIX:BeamUse} and the STAR Collaboration \cite{STAR:BeamUse}.  Much of the work in the literature has centered on effects from the transversely polarized proton~\cite{Qiu:1991wg,Kouvaris:2006zy,Ji:2006vf,Koike:2006qv,Kanazawa:2011er,Gamberg:2012iq,Kanazawa:2012kt,Gamberg:2013kla,Koike:2011nx}.  Like with the pion SSA, these include both $qgq$ and tri-gluon SGPs as well as SFPs.  The tri-gluon SGPs were shown to be negligible in the forward region~\cite{Koike:2011nx}.  The $qgq$ SGP and SFP terms are given, respectively, by~\cite{Qiu:1991wg,Kouvaris:2006zy,Ji:2006vf,Koike:2006qv,Kanazawa:2011er,Gamberg:2012iq,Kanazawa:2012kt,Gamberg:2013kla}
\begin{align}
 \frac{E_\gamma d\sigma_{(T)}^{SGP_{qgq}}(S_P)}{d^3\vec{q}} &= -\frac{4\alpha_{em} \alpha_s M}{S}
 \epsilon^{P'\!PqS_P} \int_0^1\!
dx' \int_0^1\!dx\, \delta (\hat{s}+\hat{t}+\hat{u}) \nonumber\\
 & \hspace{-2.5cm}\times \sum_a e_a^2 \frac{1}  {N_cC_F} \frac{\pi}{\hat{s}\hat{u}}
 \! \left[- \frac{1}{2N_c} f_1^{\bar{a}}(x') \,S_{\bar{a}a} + \frac{N_c}{2}
  f_1^g(x') \,S_{ga} \right]\! \!\left[F_{FT}^a(x,x)-x\frac{dF_{FT}^a(x,x)} {dx}\right], \label{e:ceSGP}\\
 \frac{E_\gamma d \sigma_{(T)}^{SFP}(S_P)}{d^3 \vec{q}} &=
-\frac{4\alpha_{em}\alpha_s M}{S} \epsilon^{P'\!PqS_P} \int_0^1\!
dx' \int_0^1\!dx\, \delta (\hat{s}+\hat{t}+\hat{u}) \nonumber\\
&\hspace{0.3cm}\times \frac{\pi} {\hat{s}}\frac{1} {2N_c} \sum_a
\bigg\{ \sum_{b} e_a e_{b} \,S^{\rm SFP}_{ab} \left[ F_{FT}^a
(0,x) - G_{FT}^a (0,x) \right] f_1^{b} (x')  \nonumber\\
&\hspace{0.3cm} \quad + \sum_{b} e_a e_{b}\, S^{\rm SFP}_{a\bar{b}}\left[ F_{FT}^a
(0,x) - G_{FT}^a (0,x) \right] f_1^{\bar{b}} (x') \nonumber\\
&\hspace{0.3cm} \qquad +\,  e_a^2\, S^{\rm SFP}_{qg}\left[ F_{FT}^a
(0,x) - G_{FT}^a (0,x) \right] f_1^g(x') \bigg\} , \label{e:ceSFP}
\end{align}
where $N_c = 3$, $C_F = 4/3$, $\alpha_{em} = e^2/4\pi$ and the hard factors $S_{\bar{a}a}$, $S_{ga}$, $S^{\rm SFP}_{ab}$, $S^{\rm SFP}_{a\bar{b}}$, $S^{\rm SFP}_{ag}$ are given in Ref.~\refcite{Kanazawa:2014nea}.  The sum for $a$ is over all quark and antiquark flavors ($a = u, \,d,\, s,\, \bar{u},\, \bar{d},\, \bar{s}$), and $\sum_{b}$ is understood to mean that the sum for $b$ is restricted over quark flavors when $a$ is a quark and over antiquark flavors when $a$ is an antiquark.

Only recently was the piece from the unpolarized proton calculated~\cite{Kanazawa:2014nea}.  This part involves the chiral-odd $qgq$ SGP function $H_{FU}(x,x)$\footnote{We note that tri-gluon correlators are only relevant for transversely polarized hadrons.} that couples to the transversity $h_1$.\footnote{In principle, one can also have a SFP term that involves $H_{FU}(0,x)$, but the hard factor vanishes when one sums overs all diagrams~\cite{Kanazawa:2014nea}. }   
The final result is given by~\cite{Kanazawa:2014nea} 
\begin{align} \label{e:co_result}
 \frac{E_\gamma d\sigma_{(U)}(S_P)}{d^3\vec{q}} &=
  -\frac{4\alpha_{em}\alpha_s M}{S} \epsilon^{P'\!PqS_P} \int_0^1
  \!dx' \int_0^1\! dx\, \delta(\hat{s}+\hat{t}+\hat{u}) \nonumber\\
 & \hspace{-1cm}\times \frac{\pi} {\hat{s}} \sum_a e_a^2 \Bigg[\!\!\left( H_{FU}^a(x',x') - x' \frac{dH_{FU}^a(x',x')}{dx'} \right)
       h_1^{\bar{a}} (x)
  \frac{S_1^{\rm SGP}}{\hat{t}} \nonumber\\
  &\hspace{1cm} - H_{FU}^a(x',x') h_1^{\bar{a}} (x)
  \,S_2^{\rm SGP} \Bigg],
\end{align}
where the hard scattering coefficients $S_1^{SGP}$, $S_2^{SGP}$ can be found in Ref.~\refcite{Kanazawa:2014nea}.
The notation for the cross section indicates that this is the entire term for the unpolarized proton.  We note that $H_{FU}(x,x)$ has a model-independent relation to the TMD Boer-Mulders function~\cite{Boer:1997nt} $h_{1}^{\perp}(x,k_T^2)$ that enters asymmetries in SIDIS and DY.  The identity reads~\cite{Boer:2003cm}
\begin{equation}
 \pi H^q_{FU}(x,x) =  h_{1}^{\perp(1),q}(x)\big|_{SIDIS} = -h_{1}^{\perp(1),q}(x)\big|_{DY}\,,
 \label{e:HFU_BM}
\end{equation}
where 
\begin{equation}
h_{1}^{\perp(1),q}(x)\equiv\!\int\!d^2k_T \,\frac{\vec{k}_T^2} {2M^2} h_{1}^\perp(x,k_T^2)\,.
\label{e:first_mom_BM}
\end{equation}

\begin{figure}[t]
 \centering \includegraphics[scale=0.25]{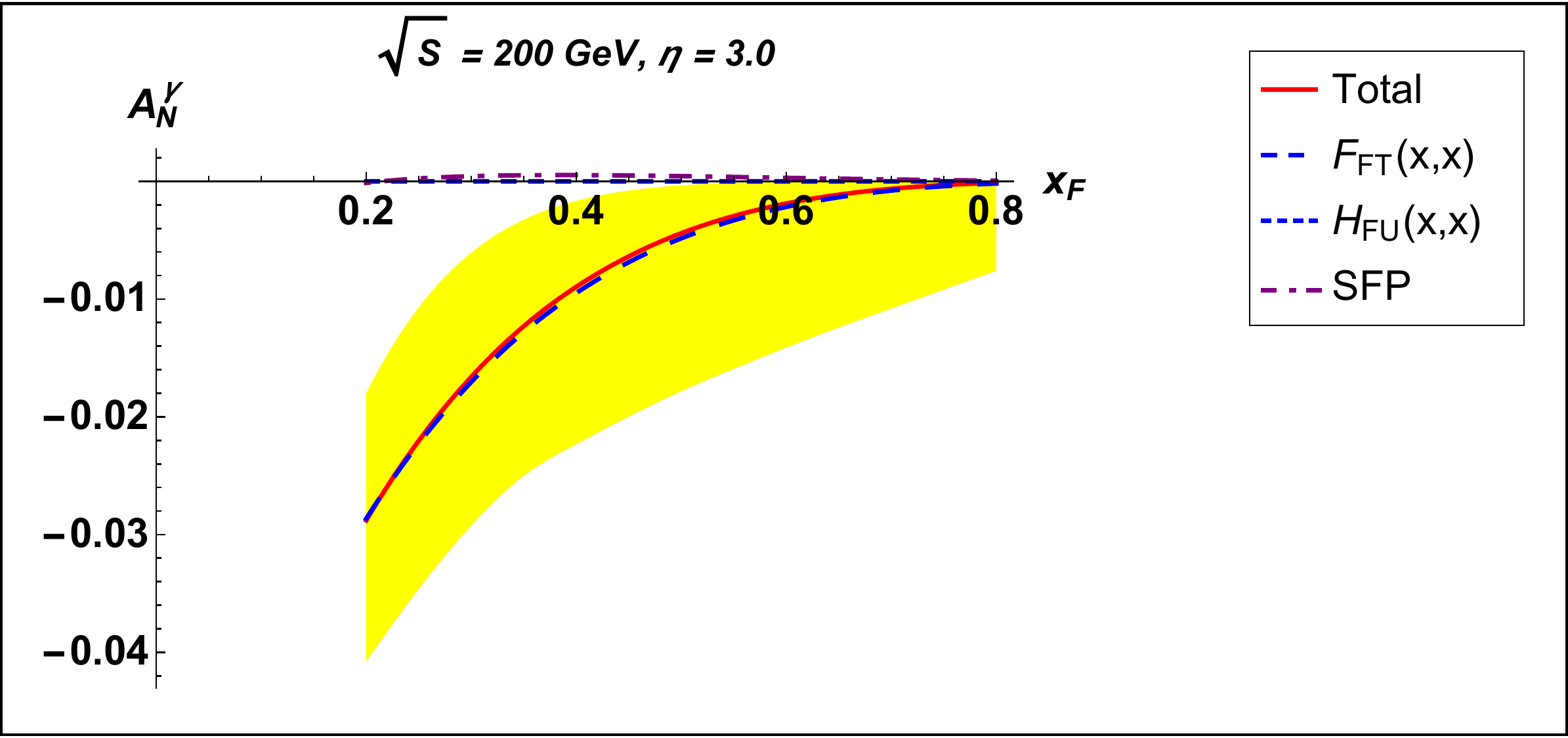}
  \includegraphics[scale=0.355]{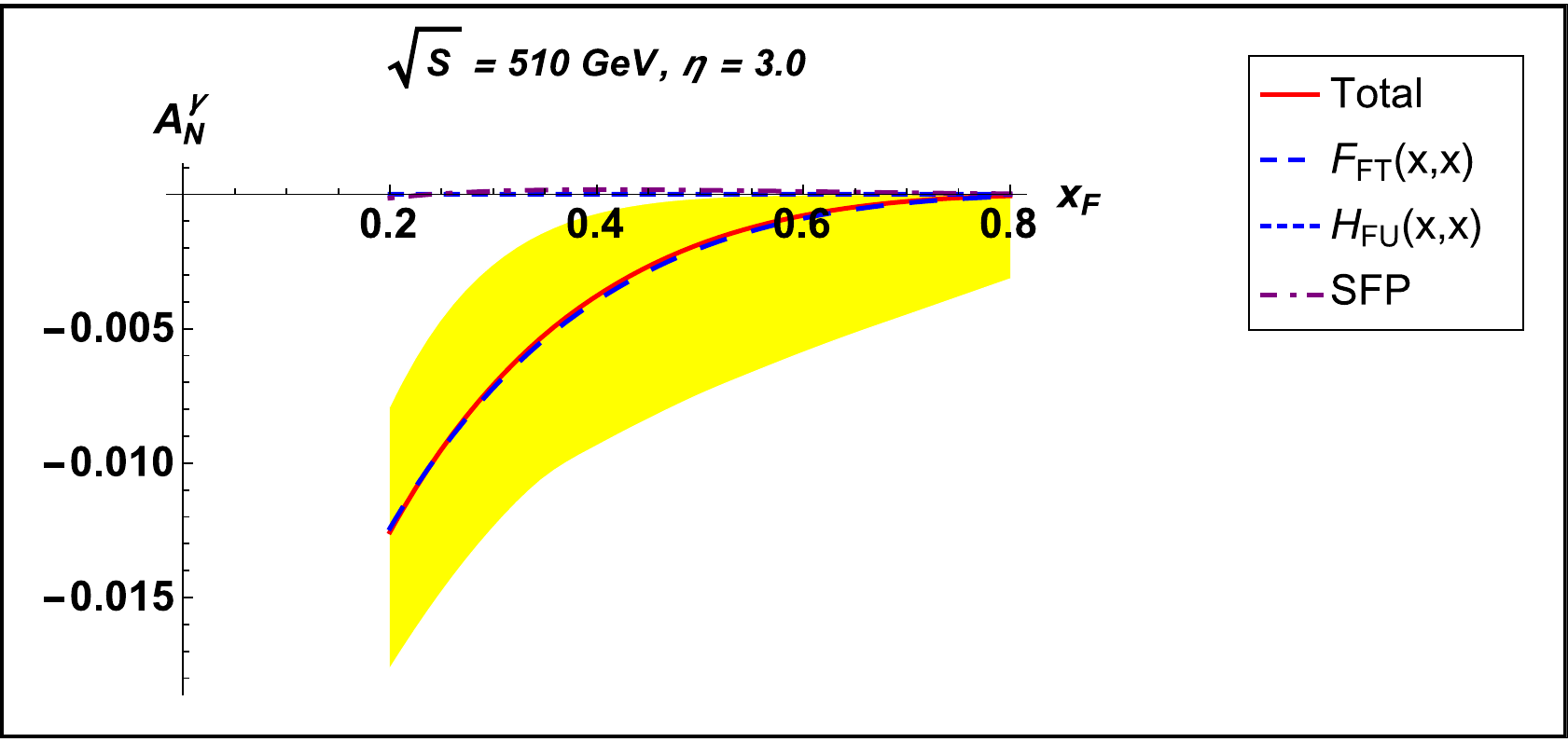}
 \caption{$A_N$ vs.~$x_F$ in direct photon production at fixed $\eta = 3.0$ and $\sqrt{S} = 200,\,510\,{\rm GeV}$.  The SGP pieces are given by the long-dashed curve for $F_{FT}(x,x)$ and the short-dashed curve for $H_{FU}(x,x)$.  The transversely polarized SFP part is shown by the dot-dashed curve.  Note again that the unpolarized SFP term vanishes.  The sum of all contributions is the solid curve.  The shaded area gives the error band in the calculation as described in Ref.~\cite{Kanazawa:2014nea}.  Figure adapted with permission from Ref.~\cite{Kanazawa:2014nea}. Copyrighted by the American Physical Society. \label{f:fixeta}}
 \vspace{-0.2cm}
\end{figure}

As with the jet asymmetry, one still cannot isolate the QS function in Eq.~(\ref{e:ceSGP}) unless the other terms (\ref{e:ceSFP}), (\ref{e:co_result}) are negligible.  Therefore, the authors of Ref.~\refcite{Kanazawa:2014nea} gave a numerical estimate for the direct photon SSA from the sum of Eqs.~(\ref{e:ceSGP}), (\ref{e:ceSFP}), and (\ref{e:co_result}) in order to determine which piece is dominant.  The functions $F_{FT}(x,x)$ and $H_{FU}(x,x)$ and were fixed by Eqs.~(\ref{e:QS_Siv}) and (\ref{e:HFU_BM}), respectively, that relate the first to the Sivers function and the second to the Boer-Mulders function.  Since at this point there is no information on the SFP functions $F_{FT}(0,x)$ and $G_{FT}(0,x)$, the assumption was made that $F_{FT}(0,x)-G_{FT}(0,x) = F_{FT}(x,x)$~\cite{Koike:2009ge}.  Based on these inputs, one finds an $A_N$ for direct photon production as given in Fig.~\ref{f:fixeta}, where the effects could be on the order of $1\%-4\%$ in magnitude.  From these plots, one sees that the QS function is the main cause the asymmetry.  Thus, data from the direct photon SSA would allow for a clean extraction of the QS function.  Moreover, since these results use Eq.~(\ref{e:QS_Siv}) to fix the $F_{FT}(x,x)$ in terms of the Sivers function, a clear signal of a negative asymmetry would be strong indication on the process dependence of the Sivers function.

\subsection{$A_{LT}$ in $p^\uparrow \!\!\vec{p}\to \{\pi,jet,\gamma\}\,X$}
The SSAs discussed so far in this section are na\"{i}ve T-odd, twist-3 observables.  As we have seen, these processes are intimately connected to the quark-gluon structure of hadrons.  In addition, there is another set of reactions that offer a complimentary, yet equally important, insight into multi-parton correlations.  These are the na\"{i}ve T-even, twist-3 longitudinal-transverse DSAs, denoted $A_{LT}$.  The most famous twist-3 DSA is $A_{LT}$ in inclusive DIS, which can be written in terms of a single function, $g_T(x)$, and has been studied  experimentally quite a bit (see Ref.~\refcite{Posik:2014usi} for the most recent data).  However, the landscape of $A_{LT}$ observables goes well beyond this simple process.  Extensive work has been performed on longitudinal-transverse DSAs in the Drell-Yan process~\cite{Jaffe:1991kp,Tangerman:1994bb,Koike:2008du,Lu:2011th}; in inclusive lepton production from $W$-boson decay in proton-proton scattering~\cite{Metz:2010xs}; for jet production~\cite{Kang:2011jw} and pion production~\cite{Kanazawa:2014tda} in lepton-nucleon collisions; and for direct photon production~\cite{Liang:2012rb}, jet/pion production~\cite{Metz:2012fq,Koike:2015yza,Koike:2016ura}, and $D$-meson production~\cite{Hatta:2013wsa} in proton-proton collisions.

Here we focus on the results for the process
\begin{equation}
p(P,S_P) + p(P',\Lambda) \rightarrow \pi(P_h) + X\,, 
\end{equation}
where all three terms in Eq.~(\ref{e:collfac}) enter.  Specifically, one receives twist-3 contributions from (a) the transversely polarized proton, (b) the longitudinally polarized proton, and (c) the (unpolarized) final-state pion.  The $qgq$ piece to (a) was calculated in Ref.~\refcite{Metz:2012fq},\footnote{The tri-gluon part to (a) has not been calculated yet, but should not be significant in the forward region.} while (b) was computed in Ref.~\refcite{Koike:2016ura} and (c) in Ref.~\refcite{Koike:2015yza}.  The three results read, respectively,
\begin{eqnarray}
\frac{E_hd\sigma^{N\!P_{qgq}}_{(T)}(S_P,\Lambda)} {d^{3}\vec{P}_h} \!&=&\! \frac{2\alpha_{s}^{2}M} {S}\,\Lambda\left(P_{hT}\cdot S_P\right)\sum_{i}\sum_{a,\,b,\,c}\,\int_0^{1}\!\frac{dz} {z^{3}}\,\int_{0}^{1}\!\frac{dx'} {x'}\int_0^1\!\frac{dx} {x}\, \delta(\hat{s}+\hat{t}+\hat{u})\,\nonumber \\[0.3cm]
&&\hspace{-2cm}\times\,g_{1}^{b}(x')D_{1}^{c}(z)\,\frac{1} {\hat{m}_{i}}\Bigg\{\!\left[g_{1T}^{(1),a}(x)-x\frac{dg_{1T}^{(1),a}(x)} {dx}\right]\!S_{g_{1T}}^{i}\nonumber\\
&&\hspace{1cm}+\,\int_{-1}^1\! dx_{1}\left[G_{DT}^{a}(x,\,x_{1})\,S_{G_{DT}}^{i}-F_{DT}^{a}(x,\,x_{1})\,S_{F_{DT}}^{i}\right]\Bigg\},\nonumber\\ \label{e:sigmahadron}
\end{eqnarray}
\begin{eqnarray}
\frac{E_hd\sigma_{(L)}(S_P,\Lambda)} {d^{3}\vec{P}_h}
&=&{2\alpha_s^2M\over S}\,\Lambda\left(P_{hT}\cdot S_P\right)\sum_i\sum_{a,b,c}\int_0^1\!{dz\over z^3}\int_0^1\!{dx\over x}
\int_0^1\! dx'\,\delta(\hat{s}+\hat{t}+\hat{u}) \nonumber\\
&&\hspace{-1.5cm}
\times\, h^a_1(x)D_1^c(z)\left[h^b_1(x')\,S^i_{h_1}+h^b_L(x')\,S^i_{h_L}
+{d h_{1L}^{\perp(1),b}(x') \over dx'}\,S^i_{h^\perp_{1L}}\right],
\label{e:ALT_Lfinal}
\end{eqnarray}
\begin{align}
\frac{E_h d\sigma^{Frag}(S_P,\Lambda)} {d^3\vec{P}_h} &= -\frac{2\alpha_s^2 M_h} {S}\,\Lambda\left(P_{hT}\cdot S_P\right)\sum_i\sum_{a,b,c}\int_0^1\!\frac{dz} {z^3}\int_0^1 \!\frac{dx^\prime} {x^\prime}\!\int_0^1\!\frac{dx} {x}\,\delta(\hat{s}+\hat{t}+\hat{u})\nonumber\\
&\hspace{2cm}\times\,h_1^a(x)\,g_1^b(x^\prime)\left(\frac{1} {z} E^c(z) \,S^i_E\right). \label{e:ALT_Frag}
\end{align}
where the hard factors $S^i$ can be found in the aforementioned references.  In Eq.~(\ref{e:sigmahadron}), $\hat{m}_i$ is a Mandelstam variable specific to each channel and can be found in Table 1 of Ref.~\refcite{Metz:2012fq}.  We have chosen to write (\ref{e:sigmahadron}) in terms of D-type functions in order to obtain a ``compact'' form for the result.\cite{Metz:2012fq}  Notice that since DSAs are na\"{i}ve T-even effects, one finds this term contains the non-pole ($N\!P$) pieces of $F_{DT}$ and $G_{DT}$.

There are several important reasons to study $A_{LT}$ in proton-proton collisions.  First, as alluded to above, one sees from Eq.~(\ref{e:sigmahadron}) that $A_{LT}$ is sensitive to the off-diagonal (i.e., $x\neq x_1$) parts of the transversely polarized $qgq$ PDFs.  This is crucial information in its own right, but one also must know the off-diagonal pieces of $F_{FT}$ and $G_{FT}$ in order to fully determine the evolution of $F_{FT}(x,x)$, \cite{Kang:2008ey, Braun:2009mi, Vogelsang:2009pj, Zhou:2008mz, Ma:2012xn} which of course is needed in SSAs.  Second, we have discussed how the pion SSA could receive a dominant contribution from the twist-3 FF $\hat{H}_{FU}^{\Im}$,~\cite{Kanazawa:2014dca} which is the imaginary part of the correlator $\hat{H}_{FU}$.  In the $A_{LT}$ case, one becomes sensitive (see Eq.~(\ref{e:ALT_Frag})) to the {\it real} part of that same correlator since $E(z)=-2z\int_z^\infty d(1/z_1)\,\hat{H}_{FU}^{\Re}(z,z_1)/(1/z-1/z_1)$~\cite{Kanazawa:2014tda,Koike:2015yza}.  Therefore, one can learn information about $\hat{H}_{FU}$ from DSAs that compliments that from SSAs.  Lastly, given the fact that there is still not a definitive conclusion as to what causes the pion SSAs in proton-proton collisions, it is crucial to test the collinear twist-3 mechanism used to describe them by measuring other observables like $A_{LT}$.  

Nevertheless, the longitudinal-transverse DSA in proton-proton collisions has not received much attention from experiments.  However, two related observables, $A_{LT}$ in SIDIS~\cite{Huang:2011bc} and in $\vec{\ell}\,n^\uparrow\to \pi\,X$~\cite{Zhao:2015wva}, have been measured and nonzero effects have been found.  Still, RHIC, with the only source of (independently manipulated) polarized proton beams in the world, has yet to explore $A_{LT}$ in $p^\uparrow \vec{p}\to \pi\,X$ despite measuring asymmetries for every other combination of proton spins.  It will be important to better develop the phenomenology of $A_{LT}$ in proton-proton collisions, which so far has been limited by the scarce information available on the relevant non-perturbative inputs, in order to motivate a measurement of this observable at RHIC.

\section{$A_N$ in lepton-nucleon collisions} \label{s:ANlN}

As we have seen, SSAs (and DSAs) in proton-proton collisions involve various pieces that are difficult to isolate.  Moreover, there is still not a definitive conclusion as to what causes these asymmetries.  Therefore, much work in the last several years has been devoted to SSAs (and DSAs) in the simpler process of $\ell\,N \to C\,X$ for various spin configurations and final states~\cite{Metz:2012ui,TPE1,TPE2,TPE5,Kang:2011jw,Anselmino:2014eza,Gamberg:2014eia,Kanazawa:2014tda,Kanazawa:2015jxa,Kanazawa:2015ajw}.  In this section, we focus on SSAs in $\ell\,N^\uparrow\to \pi\,X$.  There are several reasons this reaction has been analyzed: first, one replaces the unpolarized proton in $p^\uparrow p\to \pi\,X$ with a lepton, which drastically reduces the number of Feynman diagrams in the calculation; second, one eliminates (at leading order (LO)) tri-gluon correlators and SFPs in the transversely polarized proton; third, HERMES~\cite{Airapetian:2013bim} and Jefferson Lab (JLab) Hall A~\cite{Allada:2013nsw} have measured this observable, and a future Electron-Ion Collider (EIC)~\cite{Accardi:2012qut} will also provide valuable data on this reaction.

The analytical result for the process
\begin{equation}
\ell(l) + N(P,S_N) \rightarrow \pi(P_h)+X
\end{equation}
is similar in structure to Eqs.~(\ref{finalcr}), (\ref{e:sigmaFrag}) and reads~\cite{Gamberg:2014eia}
\begin{eqnarray}
 \frac{ E_h d\sigma(S_N)} {d^3\vec{P}_h} & = & - \frac{8\alpha_{\rm em}^2} {S} \,
\epsilon^{\l P P_h S_N} \,
\sum_a e_a^2 \int_{0}^1\! \frac{dz} {z^3}\,\int_0^1\!\frac{dx} {x}\, \delta(\hat{s} + \hat{t}+\hat{u})
\nonumber \\
& & \times\, \Bigg\{\!\!-\!\frac{\pi M} {\hat{u}}\,D_1^{a}(z) \bigg(F_{FT}^a(x,x)-x\frac{dF_{FT}^a(x,x)} {dx}\bigg)\!\bigg[\frac{\hat{s}(\hat{
s}^2+\hat{u}^2)} {2\hat{t}^{3}}\bigg]
\nonumber \\
& & \hspace{0.5cm}+\,\frac{M_h} {-x\hat{u}-\hat{t}}\,\,h_{1}^{a}(x)\,\Bigg\{\!\!\bigg(H_1^{\perp(1),a}(z)-z\frac{dH_1^{\perp(1),a}(z)} {dz}\bigg)\!\bigg[\frac{(1-x)\hat{s}\hat{u}} {\hat{t}^{2}}\bigg] 
\nonumber \\
& & \hspace{-1.5cm} +\, \frac{1} {z} \, H^{a}(z) \bigg[ \frac{\hat{s} (\hat{s}^2 +(x-1)\hat{u}^2)} {\hat{t}^{3}}\bigg] 
+ \frac{2} {z} \! \int_z^\infty \! \frac{dz_1} {z_1^2} \, \frac{1} {\left(\frac{1} {z} -\frac{1} {z_{1}}\right)^{\!2}} \, \hat{H}_{FU}^{a,\Im}(z,z_{1}) 
\bigg[ \frac{x\hat{s}^2\hat{u}} {\hat{t}^{3}} \bigg]\!\Bigg\}\!\Bigg\} \,,
\nonumber\\ \label{e:lNhXUT}
\end{eqnarray}
where the Mandelstam variables are defined like in Eqs.~(\ref{e:Mand}), (\ref{e:pMand}) but with $P'\to l$ and $x' = 1$.

\begin{figure*}[t]
\centering
      \includegraphics[scale=0.332]{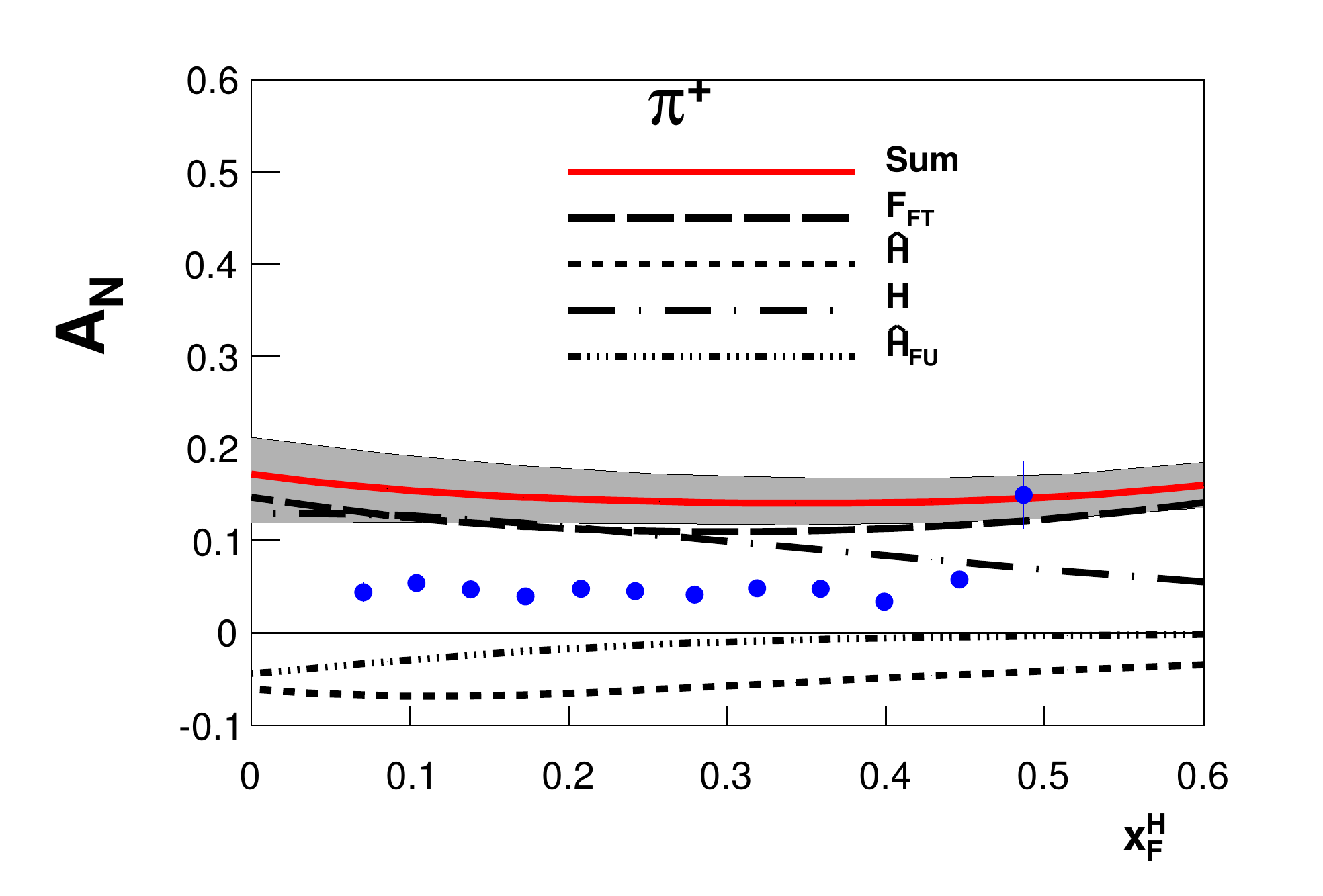}\hspace{-0.7cm}
      \includegraphics[scale=0.332]{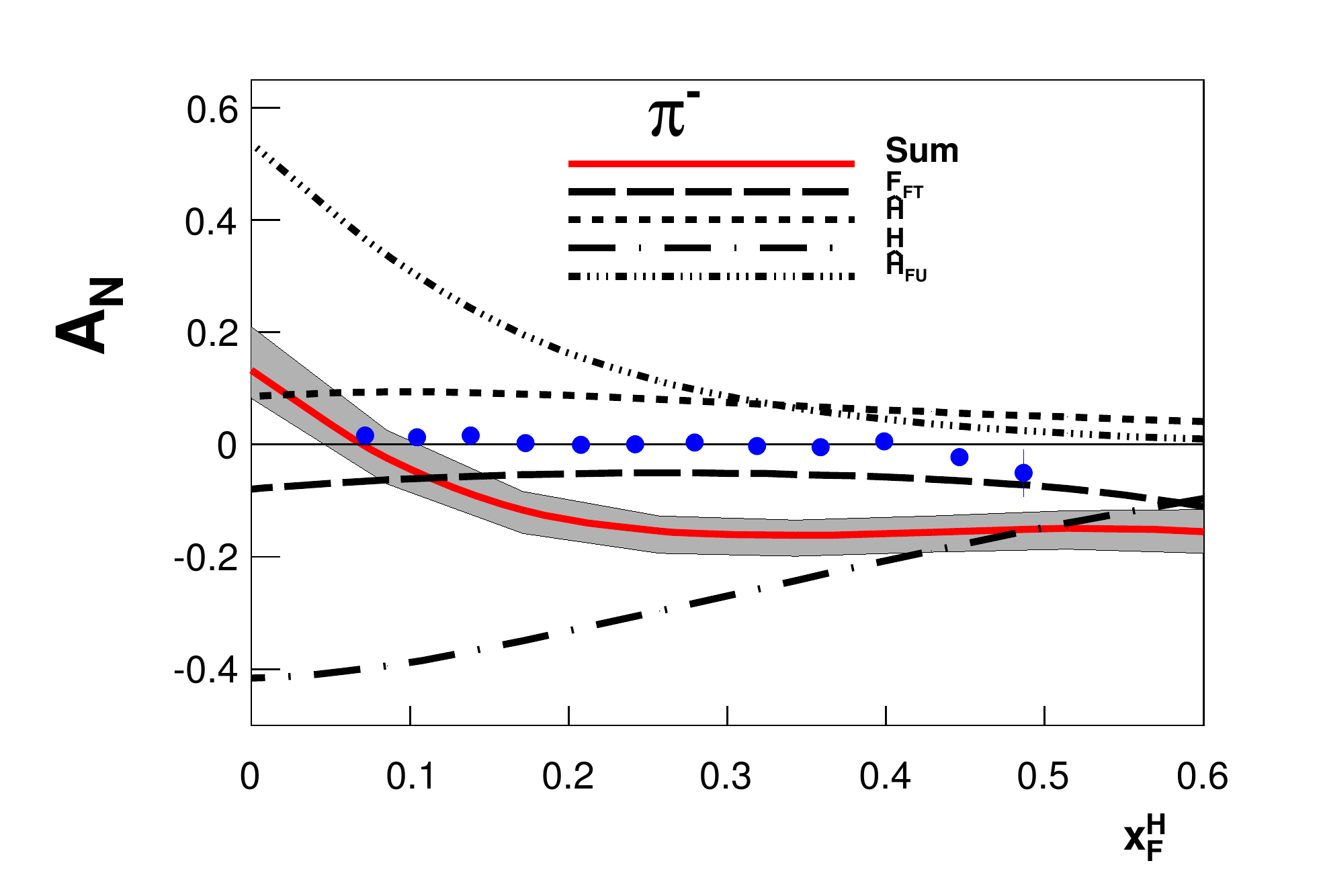}
 \vspace{-0.75cm}
\caption{$A_N$ vs.~$x_F^H$ for charged pion production in electron-proton collisions at HERMES kinematics ($P_{hT} = 1 \, \rm{GeV}$ and $\sqrt{S} = 7.25 \, \rm{GeV}$). 
The data are from Ref.~\cite{Airapetian:2013bim}. 
The contributions from $F_{FT}(x,x)$, $\hat{H}(z)\equiv H_1^{\perp(1)}(z)$, $H(z)$, and $\hat{H}^{\Im}_{FU}(z,z_1)$ in Eq.~(\ref{e:lNhXUT}) are, respectively, the dashed line, dotted line, dot-dashed line, and 3-dotted-dashed line. The solid curve is the total contribution from all terms.
The error band is due to uncertainties in the Sivers, Collins, and transversity functions estimated in Refs.~\cite{Anselmino:2008sga,Anselmino:2013vqa}.  
Note that positive $x_F^{\rm H}$ corresponds to pions in the backward direction with respect to the target proton.  Figure reprinted with permission from \href{http://dx.doi.org/10.1103/PhysRevD.90.074012} {L.~Gamberg, et al., Phys.~Rev.~D90, 074012 (2014)}.  Copyright (2014) by the American Physical Society.}
\label{fig:an_hermes_pipm_xf}
\end{figure*}

Based on Eq.~(\ref{e:lNhXUT}), one can give an estimate for this observable by using known inputs for the non-perturbative functions.  Specifically, the authors of Ref.~\refcite{Gamberg:2014eia} used Eq.~(\ref{e:QS_Siv}) to fix $F_{FT}(x,x)$ in terms of the Sivers function, Eq.~(\ref{e:H1perp}) to fix $H_1^{\perp(1)}(z)$ in terms of the Collins function, and the extraction of $\hat{H}^{\Im}_{FU}(z,z_1)$ from Ref.~\refcite{Kanazawa:2014dca} (along with the QCD EOM relation (\ref{e:EOM_FF})) to fix $\hat{H}^{\Im}_{FU}(z,z_1)$ and $H(z)$.  Their results for HERMES kinematics\footnote{Note that in the HERMES convention, positive Feynman-$x$ (which is denoted by $x_F^H$) corresponds to hadrons going in the direction of the lepton, i.e., in the {\it backward} region of the target proton. 
This convention has the opposite sign compared to $x_F$ used in the proton-proton collisions, i.e., $x_F^H = -x_F$.}  are shown in Fig.~\ref{fig:an_hermes_pipm_xf}.  One sees that the phenomenological curves generally disagree with the data, although most likely the theoretical error band is underestimated (see Ref.~\refcite{Gamberg:2014eia} for details).  Moreover, due to the typically low value of $P_{hT}$ at HERMES ($\sim 1-2\,{\rm GeV}$), and the fact that almost all the pions come from quasi-real photoproduction,~\cite{Airapetian:2013bim} the majority of the data from HERMES (and JLab\footnote{In fact, the JLab measurements are in the non-perturbative regime ($P_{hT}<1\, {\rm GeV}$) and, therefore, one cannot rigorously compare their data to theory estimates.}) is not covered by a LO calculation.  It will be important in the future to conduct a next-to-leading order (NLO) computation of $A_N$ for this process.  Already a NLO calculation of the unpolarized cross section has been performed that shows quite large corrections to the LO formalism~\cite{Hinderer:2015hra,Schlegel:2015hga}.

\begin{figure*}[t]
\centering
      \includegraphics[scale=0.325]{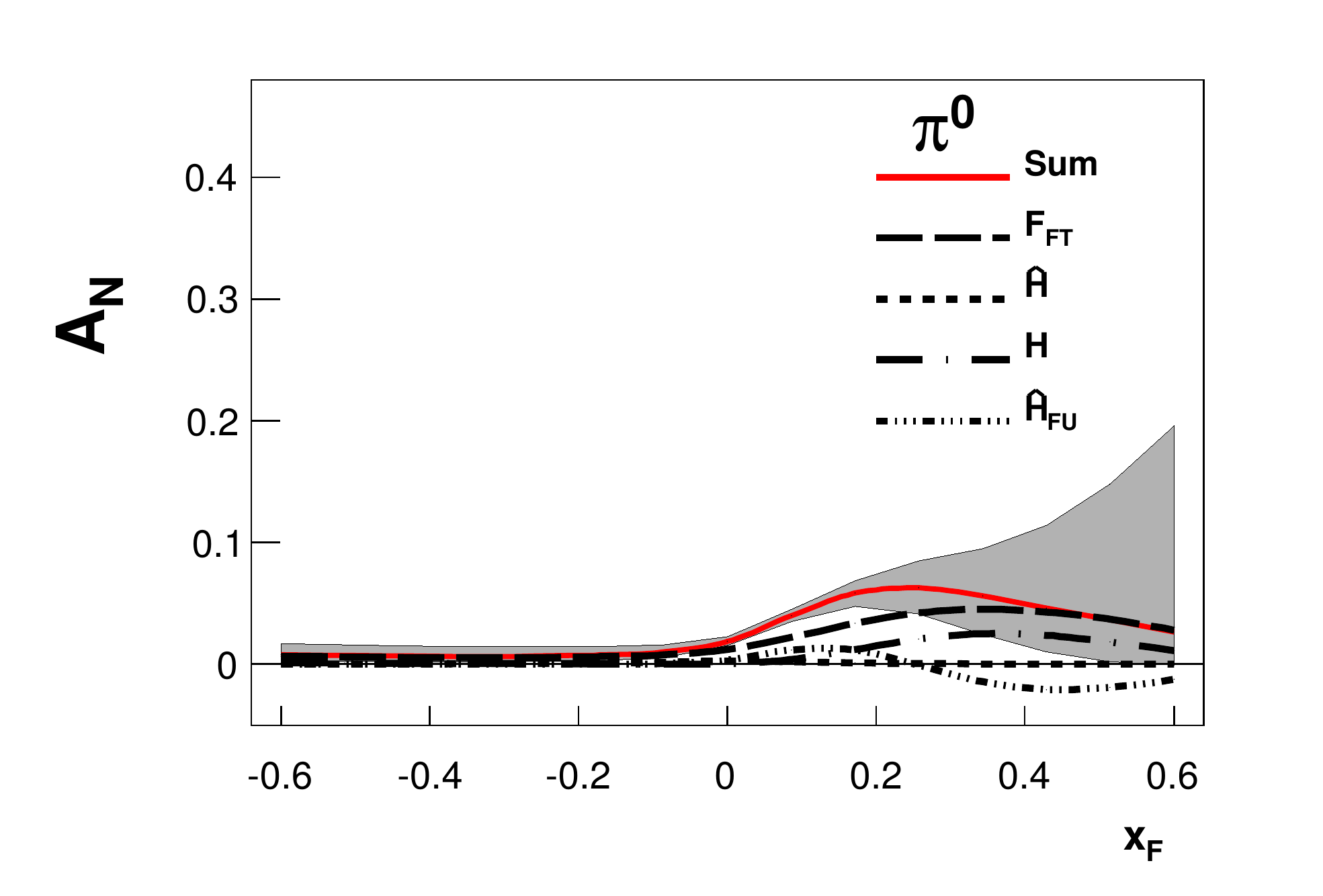} \hspace{-0.7cm}
      \includegraphics[scale=0.325]{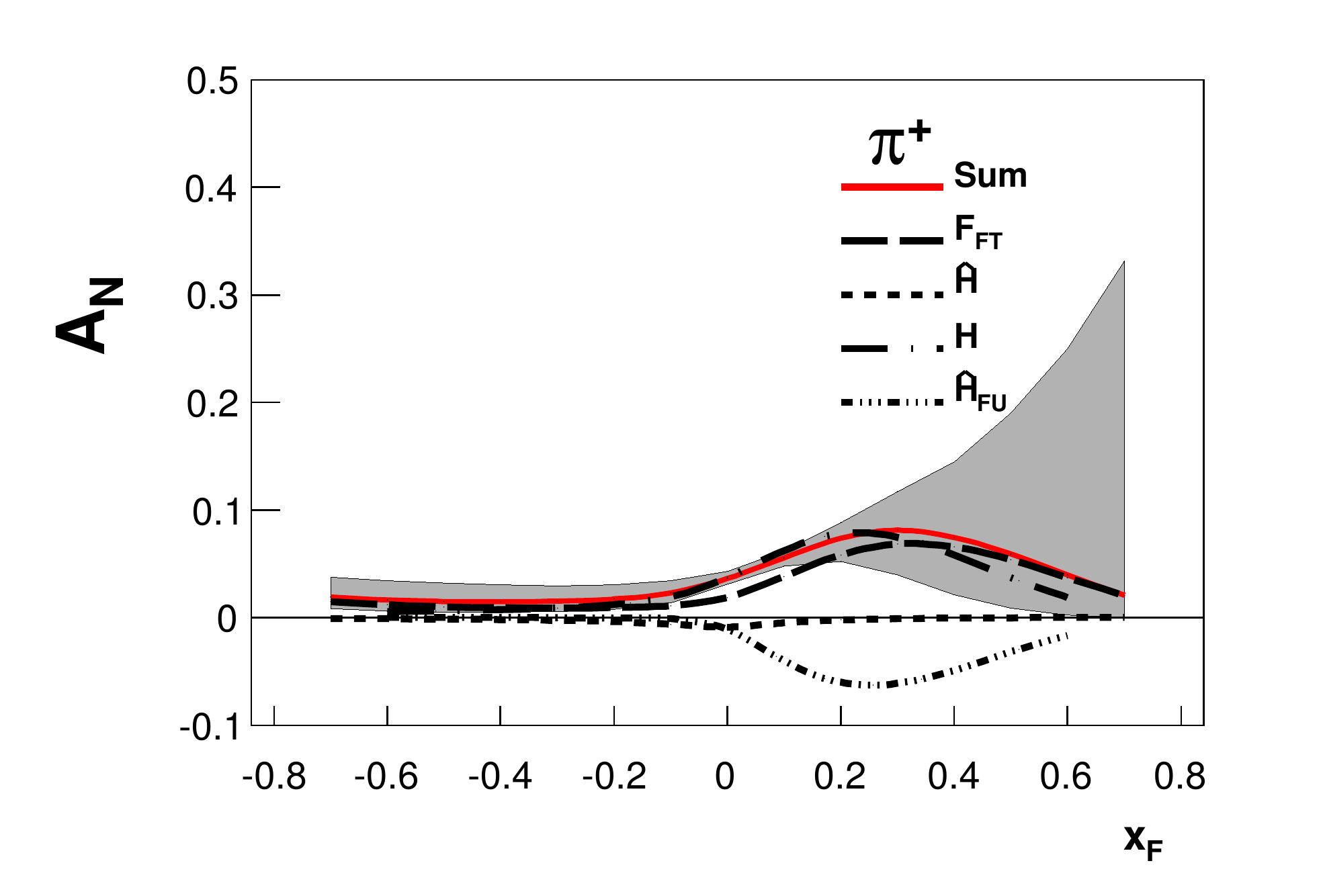}
   \vspace{-0.3cm}
\caption{$A_N$ vs.~$x_F$ for $\pi^0$ and $\pi^+$ production in electron-proton collisions at EIC kinematics ($P_{hT} = 3 \; \rm{GeV}$ and $\sqrt{S} = 63 \; \rm{GeV}$). 
The lines are the same as in Fig~\ref{fig:an_hermes_pipm_xf}.  Figure reprinted with permission from \href{http://dx.doi.org/10.1103/PhysRevD.90.074012} {L.~Gamberg, et al., Phys.~Rev.~D90, 074012 (2014)}.  Copyright (2014) by the American Physical Society.}
\label{fig:an_eic_pi0_xf}
\end{figure*}

\begin{figure*}[t]
\centering
      \includegraphics[scale=0.332]{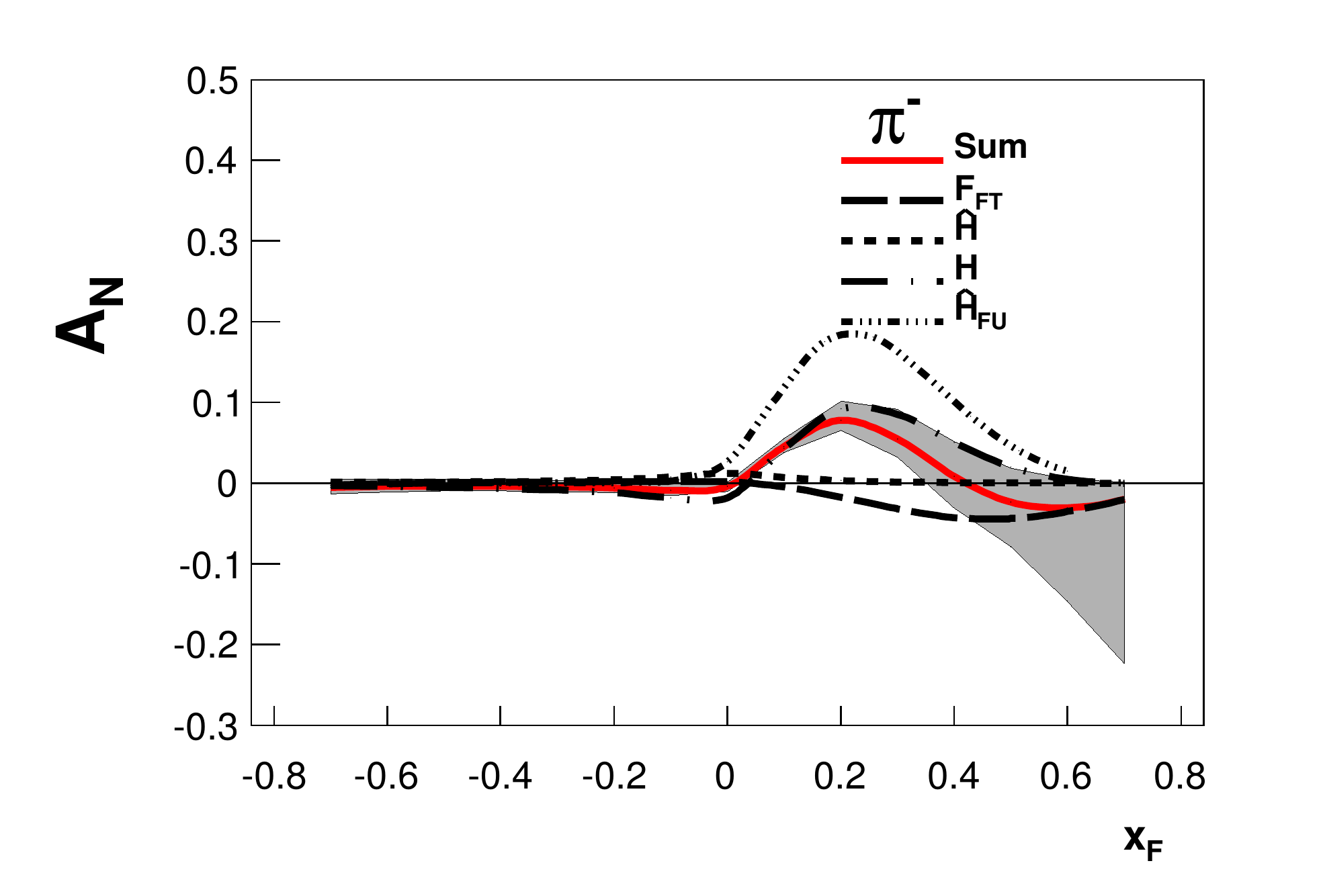}\hspace{-0.7cm}
      \includegraphics[scale=0.332]{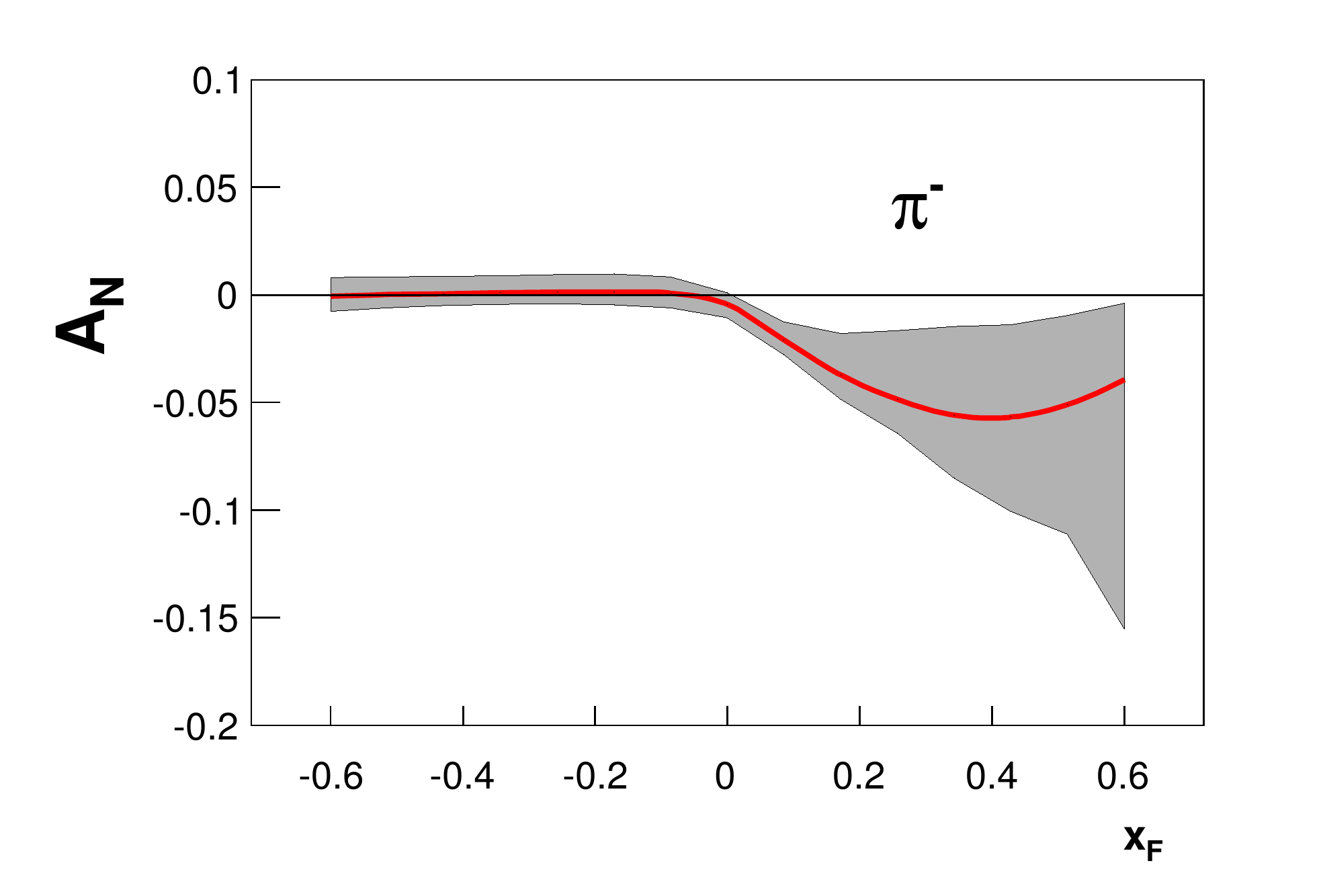}
 \vspace{-0.8cm}
\caption{$A_N$ vs.~$x_F$ for $\pi^-$ production in electron-proton collisions at EIC kinematics ($P_{hT} = 3 \; \rm{GeV}$ and $\sqrt{S} = 63 \; \rm{GeV}$).  The plots show the effect of  $\hat{H}_{FU}^{\Im}$ on the asymmetry, with the left one having $\hat{H}_{FU}^{\Im} \neq 0$ and the right one having $\hat{H}_{FU}^{\Im} = 0$.
The lines are the same as in Fig~\ref{fig:an_hermes_pipm_xf}.  Figure reprinted with permission from \href{http://dx.doi.org/10.1103/PhysRevD.90.074012} {L.~Gamberg, et al., Phys.~Rev.~D90, 074012 (2014)}.  Copyright (2014) by the American Physical Society.}
\label{fig:an_eic_pipm_xf}
\end{figure*}

Beyond the current data we have on this observable, a future EIC can offer unique insight into $A_N$ in lepton-nucleon collisions.  Most notably, an EIC will be able to measure this process in the {\it forward} region of the proton, analogous to what has already been done at RHIC.  An estimate for this asymmetry is given in Figs.~\ref{fig:an_eic_pi0_xf}, \ref{fig:an_eic_pipm_xf}, and it could be as large as $\sim 20\%-30\%$.  In addition, an EIC would be able to test the recently proposed mechanism behind $A_N$ in proton-proton collisions, i.e., that these effects are due to the twist-3 FF $\hat{H}_{FU}^{\Im}$~\cite{Kanazawa:2014dca}.  For example, in Fig.~\ref{fig:an_eic_pipm_xf}, one sees that $\hat{H}_{FU}^{\Im}$ can cause $A_N$ for $\pi^-$ to change sign.  This demonstrates how one can learn about SSAs in proton-proton collision by studying the analogous observable in lepton-nucleon collisions.

\section{Recent theoretical progress in the twist-3 formalism} \label{s:progress}
There has been some progress made recently in solidifying the twist-3 formalism, most notably from Ref.~\refcite{Kanazawa:2015ajw}, which we summarize here.  As one might gather from the previous sections, the common ingredient to any twist-3 observable is that there must be a transverse vector in the process, e.g., a proton carrying a transverse spin.  Two lightcone vectors are needed in order to define a transverse direction relative to this proton.  One of these vectors is naturally the momentum $P$ of the proton (assuming its mass is negligible) and the other vector $n$ must be ``adjoint'' to $P$.  That is, $n$ must satisfy
\begin{equation}
n^2=0,\quad P\cdot n = 1\,. \label{e:Pn}
\end{equation}
From this, one can define a transverse vector relative to $P,n$ as $a_T^\mu = a^\mu - (a\cdot P)\,n^\mu - (a\cdot n)\,P^\mu$.  Likewise, there are also lightcone vectors $P_h,m$ associated with the final-state hadron that satisfy
\begin{equation}
m^2=0,\quad P_h\cdot m = 1\,, \label{e:Phm}
\end{equation}
with a transverse vector relative to $P_h,m$ defined as $a_\perp^\mu = a^\mu - (a\cdot P_h)\,m^\mu - (a\cdot m)\,P_h^\mu$.  Thus, transverse spin observables, {\it a priori}, depend on $n$ and $m$. 

Therefore, since the conditions (\ref{e:Pn}), (\ref{e:Phm}) do not completely fix $n$ and $m$,~\cite{Kanazawa:2015ajw} twist-3 cross sections  could depend on the choice one makes for these vectors, which would violate Lorentz invariance~\cite{Kanazawa:2015ajw}. In fact, one does find that twist-3 observables calculated in different frames appear to give different results~\cite{Kanazawa:2014tda,Kanazawa:2015ajw}.  Nevertheless, with the use of so-called Lorentz invariance relations (LIRs), which are given in \ref{a:LIR}, one can explicitly eliminate this spurious dependence on $n$ and $m$ and bring twist-3 cross sections into a manifestly Lorentz invariant form.~\cite{Kanazawa:2015ajw}  
\begin{table}[h]
\tbl{The reduction of Table~\ref{t:T3func} to the set of fundamental (independent) collinear twist-3 functions.}
{\centering \includegraphics[scale=0.45]{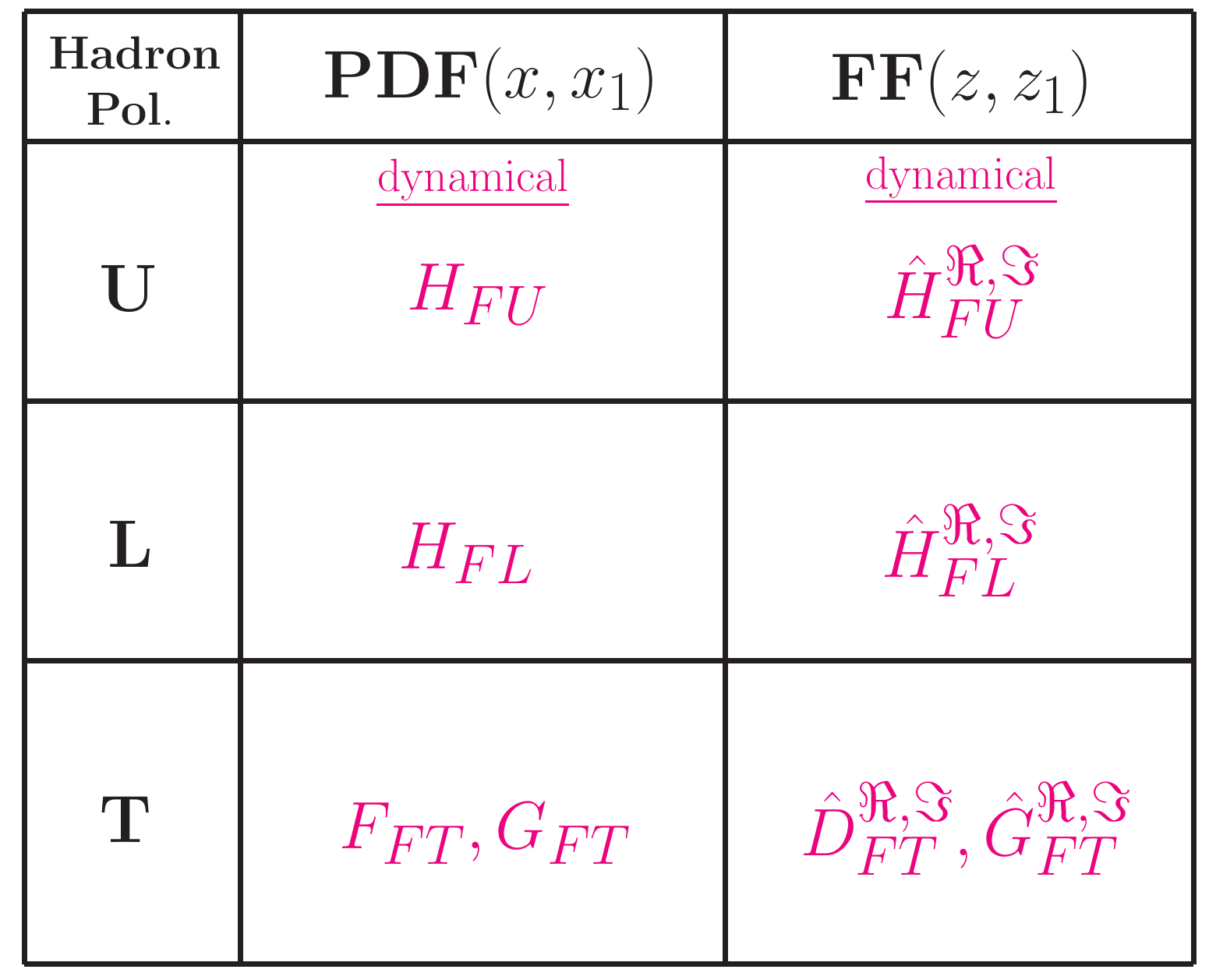}} \label{t:T3func_red}
\end{table}
The authors of Ref.~\refcite{Kanazawa:2015ajw} demonstrated this for asymmetries in $\ell\,N\to h\,X$ for various lepton/nucleon/hadron polarizations:~$A_{UTU}\equiv A_N$, $A_{LTU}\equiv A_{LT}$, $A_{UUT}$, $A_{LUT}$.  Furthermore, agreement was found between calculations in two different color gauges, and the cross sections were also shown to satisfy electromagnetic gauge invariance.~\cite{Kanazawa:2015ajw}  Therefore, the analytical techniques use to calculate twist-3 observables in proton-proton and lepton-nucleon collisions seem to be on solid ground.

In addition, given the full set of EOM relations and LIRs that connect intrinsic, kinematical, and dynamical twist-3 functions (see \ref{a:EOM} and \ref{a:LIR}), one can actually write all intrinsic and kinematical functions in terms of dynamical ones.~\cite{Kanazawa:2015ajw}  This is shown explicitly in \ref{a:dyn}, which allows us to reduce Table~\ref{t:T3func} to Table~\ref{t:T3func_red}.  Thus, observables involving the transverse spin of hadrons give us direct access to multi-parton correlations, and these are the fundamental (independent) functions that drive twist-3 observables.

\section{Outlook and future research} \label{s:concl}
Transverse spin observables in hadronic processes have been a fruitful source of research for 40 years.  In this review, we have focused on single-inclusive hard scattering processes where one can apply collinear factorization.  These reactions provide a critical test of the pQCD framework, and we have outlined the recent progress in this area of spin physics.  Some of the highlights include:

\begin{enumerate} [label={\arabic*)}]
\item The large SSAs seen in $p^\uparrow p\to \pi\, X$~\cite{Klem:1976ui,Bunce:1976yb,Adams:1991rw,Krueger:1998hz,
Allgower:2002qi,Adams:2003fx,Adler:2005in,Lee:2007zzh,:2008mi,:2008qb,Adamczyk:2012qj,Adamczyk:2012xd,Bland:2013pkt,Adare:2013ekj} are not due to effects in the transversely polarized proton (i.e., the QS function)~\cite{Kang:2011hk,Kang:2012xf,Metz:2012ui} but could mainly arise from parton fragmentation effects during the formation of the final-state pion~\cite{Kanazawa:2014dca}. \\

\item The SSA in $p^\uparrow p\to \gamma\, X$ can provide a clean extraction of the QS function as well as test the process dependence of the Sivers function~\cite{Gamberg:2013kla,Kanazawa:2014nea}.  Data is on tape from PHENIX and STAR for this observable.~\cite{PHENIX:BeamUse,STAR:BeamUse}\\

\item Longitudinal-transverse DSAs can provide complimentary, yet equally important, information as SSAs on the quark-gluon structure of hadrons.  The main analytical results have been calculated~\cite{Metz:2012fq,Koike:2015yza,Koike:2016ura}, although more work is needed on the phenomenology.  No experiments at RHIC have ever measured this asymmetry.\\

\item SSAs and DSAs in lepton-nucleon collisions allow one to gain insight into, as well as the test the mechanisms behind, the analogous effects in proton-proton collisions~\cite{Metz:2012ui,Kang:2011jw,Anselmino:2014eza,Gamberg:2014eia,Kanazawa:2014tda,Kanazawa:2015jxa,Kanazawa:2015ajw}.  These observables themselves have been measured at HERMES~\cite{Airapetian:2013bim} and JLab~\cite{Allada:2013nsw} in the backward region, and unique information on SSAs in the forward region will be accessible at an EIC~\cite{Gamberg:2014eia}.\\

\item Even though within the collinear twist-3 framework there are a plethora of matrix elements (intrinsic, kinematical, and dynamical) that enter into these transverse spin effects (see Table \ref{t:T3func}), the fundamental (independent) objects are the dynamical functions~\cite{Kanazawa:2015ajw}.  Thus, SSAs and DSAs give direct access to multi-parton correlations in hadrons.
\end{enumerate}

Nevertheless, even with these advances, there are still unresolved issues in our understanding of the transverse spin effects discussed here that will be addressed by future theoretical and experimental research.  First, from the theoretical side, one must determine how the additional constraints from LIRs involving twist-3 FFs~\cite{Kanazawa:2015ajw} affect the phenomenological extraction of these functions and fits to data, like those in Refs.~\refcite{Kanazawa:2014dca,Gamberg:2014eia}.  Next, (unweighted) twist-3 cross sections should be calculated at NLO, which would extend the NLO calculations of $P_{hT}$-weighted twist-3 SSAs~\cite{Vogelsang:2009pj,Dai:2014ala,Yoshida:2016tfh}.  Lepton-nucleon collisions are the most reasonable observable to perform such a computation, and a NLO result for this reaction also has practical phenomenological applications (see the discussion in Ref.~\refcite{Gamberg:2014eia}).  Already, progress has been made on this front with a NLO calculation of the unpolarized cross section~\cite{Hinderer:2015hra,Schlegel:2015hga}.  Furthermore, there has been tremendous work over the last several years on factorization and evolution in TMD processes (see Ref.~\refcite{Rogers:2015sqa} for an overview), where there are two scales $Q_1,Q_2$ with $Q_1 \ll Q_2$.  These TMD reactions are related to collinear observables in the region $Q_1\sim Q_2$, and one should further examine and extend this connection in order to improve the phenomenology in processes like SIDIS and Drell-Yan (see, e.g., Ref.~\refcite{Collins:2016hqq}).

In addition to these theoretical efforts, there are several near- and long-term experimental measurements that will aid in finding a definitive answer to what causes SSAs in proton-proton (and proton-nucleus) collisions.  First, at RHIC, along with the $A_N$ data already mentioned in this review, there will be more data upcoming from STAR on photon, $W^\pm$, $Z$, and DY final states as well as charged hadrons and ``flavor enhanced'' jets~\cite{Aschenauer:2016our}.  In order to explore whether $A_N$ is the result of a 2-to-2 hard scattering mechanism (like in the collinear twist-3 framework), there will also be measurements of SSAs in diffractive events~\cite{Aschenauer:2016our}. (We mention that other alternative mechanisms to explain SSAs have been proposed in Refs.~\citen{Hoyer:2006hu,Qian:2011ya,Kovchegov:2012ga,Troshin:2012fr}.)  Furthermore, RHIC will look to expand their data on SSAs in proton-nucleus collisions~\cite{Dilks:2016}, which has been a fruitful source of research connecting (transverse) spin and small-$x$/saturation physics~\cite{Boer:2006rj,Kang:2011ni,Kovchegov:2012ga,Kang:2012vm,Zhou:2013gsa,Altinoluk:2014oxa,Schafer:2014zea,Zhou:2015ima,Boer:2015pni,Hatta:2016wjz}.

Along with these measurements at RHIC, there is also the potential for SSAs to be examined at a polarized fixed target experiment, called AFTER, at the Large Hadron Collider (LHC)~\cite{Brodsky:2012vg,Lansberg:2014myg}.  The study of SSAs at the LHC would provide more statistics and a higher reach in $P_{hT}$ than RHIC.  This would allow AFTER to not just supplement the RHIC data, but also give new insight into outstanding questions involving SSAs~\cite{Kanazawa:2015fia,Anselmino:2015eoa} (e.g., whether or not the asymmetries fall off at large $P_{hT}$).  Finally, an EIC would give unique information on SSAs in lepton-nucleon collisions in the forward region, which are intimately connected to the analogous asymmetries in proton-proton collisions.  Thus, such measurements would allow one to test what causes SSAs proton-proton reactions and, because there will be more precise data, call for new rigorous phenomenological analyses~\cite{Kang:2011jw,Anselmino:2014eza,Gamberg:2014eia,Kanazawa:2014tda,Kanazawa:2015jxa,Kanazawa:2015ajw}.

\section*{Acknowledgments}

This work has been supported by the RIKEN BNL Research
Center.  I would like to thank Penn State University-Berks for their hospitality during the completion of this paper.\\

\appendix

\section{Operator definitions of collinear twist-3 functions} \label{a:def}
In this appendix, we give the operator definitions of the collinear twist-3 functions used in the main text, which are the same as those in Ref.~\refcite{Kanazawa:2015ajw}.  We start with the intrinsic functions, which are based on the correlators
\begin{eqnarray}
\Phi^q_{ij}(x) &=&\int_{-\infty}^{\infty}\frac{d\lambda}{2\pi}\,\mathrm{e}^{i\lambda x}\langle P,S|\,\bar{q}_j(0)\,\mathcal{W}[0\,;\,\lambda n]\, q_i(\lambda n)\,|P,S\rangle\,,\label{Phix}
\end{eqnarray}
\begin{eqnarray}
\Delta^{q}_{ij}(z) & = & \frac{1}{N_{c}}\sum_{X}\hspace{-0.5cm}\int\int_{-\infty}^{\infty}\frac{d\lambda}{2\pi}\,\mathrm{e}^{-i\frac{\lambda}{z}}\langle0|\, \mathcal{W}[ \pm \infty m\,;\,0]\,q_i(0)\,|P_{h}S_{h};\,X\rangle \nonumber\\
&& \times \langle P_{h}S_{h};\,X|\,\bar{q}_j(\lambda m)\,\mathcal{W}[\lambda m\,;\,\pm\infty m]\,|0\rangle\,,\label{Deltaz}
\end{eqnarray}
where $q$ is a quark field and $\mathcal{W}$ is a Wilson line that renders the matrix elements color gauge invariant.  The momenta $P$ is that of the nucleon and $P_h$ that of the outgoing hadron.  The number of colors is indicated by $N_c$. The two definitions (\ref{Phix}) and (\ref{Deltaz}) also use two lightcone vectors $n$ and $m$ that satisfy $n^2=0$, $P\cdot n=1$ and $m^2=0$, $P_h\cdot m=1$.  The spin $S$ of the nucleon satisfies $S^2 = -1$ and $P\cdot S = 0$, and likewise for $P_h, S_h$.

The correlators $\Phi$ and $\Delta$ can be parameterized in terms scalar functions, and (up to twist-3 accuracy) read~\cite{Mulders:1995dh,Boer:1997mf,Goeke:2005hb} 
\begin{eqnarray}
\Phi^{q}(x) & \!=\! & \frac{1}{2}\Bigg(\slash P\, f_{1}^{q}(x)+M\, e^{q}(x)
 -M\,(S\cdot n)\,\slash P\gamma_{5}\, g_{1}^{q}(x)+\tfrac{1}{2}M^{2}\,(S\cdot n)\,[\slash P,\slash n]\gamma_{5}\, h_{L}^{q}(x)\nonumber \\
 &  & -\frac{1}{2}[\slash P, 
\slash  S]\gamma_{5}\, h_{1}^{q}(x)  -M\,(\slash S-(S\cdot n)\slash P)\gamma_{5}\, g_{T}^{q}(x)\Bigg)\,,\label{eq:PhixParam}
\end{eqnarray}
\begin{eqnarray}
\Delta^{q}(z) & \!=\! & \,\frac{1}{z}\Bigg(\slash P_{h}\, D_{1}^{q}(z)+M_{h}\, E^{q}(z)+\tfrac{i}{2}M_{h}[\slash P_{h},\slash m]\, H^{q}(z) -M_{h}\,(S_{h}\cdot m)\,\slash P_{h}\gamma_{5}\, G_{1}^{q}(z)\nonumber\\
& & +\tfrac{1}{2}M_{h}^{2}(S_{h}\cdot m)\,[\slash P_{h},\slash
 m]\gamma_{5}\, H_{L}^{q}(z) - M_h^2 (S_h\cdot m) i \gamma_5 \,
 E_L^q(z) -\frac{1}{2}[\slash P_{h},
\slash S_{h}]\gamma_{5}\, H_{1}^{q}(z) \nonumber\\
&& -M_{h}\,\epsilon^{P_{h}m\alpha S_{h}}\gamma_{\alpha}\, D_{T}^{q}(z) -M_{h}\,(\slash S_{h}-(S_{h}\cdot m)\slash P_{h})\gamma_{5}\, G_{T}^{q}(z)\Bigg)\,,\nonumber \\ \label{eq:DeltazParam}
\end{eqnarray}
where the nucleon mass $M$ and hadron mass $M_h$ are introduced to make the PDFs and FFs dimensionless.  The convention for the Levi-Civita tensor is $\epsilon^{0123} =+1$.

We next turn to the kinematical functions, which are derived from the following TMD matrix elements,
\begin{eqnarray}
\Phi^q_{ij}(x,k_T)&\!=\!& \int_{-\infty}^{\infty}\frac{d\lambda}{2\pi}\int\!\frac{d^2z_T}{(2\pi)^2}\,\mathrm{e}^{ix\lambda+ik_T\cdot z_T}\langle P,S|\,\bar{q}_j(0)\,\mathcal{W}[0\,;\,\infty n]\mathcal{W}[\infty n\,;\,\infty n+\infty z_T]\nonumber\\
&&\times\,\mathcal{W}[\infty n+\infty z_T\,;\,\infty n+z_T]\mathcal{W}[\infty n+z_T\,;\,\lambda n+z_T]\,q_i(\lambda n+z_T)\,|P,S\rangle\,,\nonumber\\\label{PhiTMD}
\end{eqnarray}
\begin{eqnarray}
\Delta^q_{ij}(z,p_\perp)& \!=\!& \frac{1}{N_{c}}\sum_{X}\hspace{-0.5cm}\int\int_{-\infty}^{\infty}\frac{d\lambda}{2\pi}\int \frac{d^2 z_\perp}{(2\pi)^2}\,\mathrm{e}^{-i\frac{\lambda}{z}-ip_\perp\cdot z_\perp}\langle0|\mathcal{W}[\pm\infty m+\infty z_\perp\,;\,\pm \infty m]\nonumber\\
& &\times\, \mathcal{W}[ \pm \infty m\,;\,0]\,q_i(0)|P_{h}S_{h};\,X\rangle\langle P_{h}S_{h};\,X|\,\bar{q}_j(\lambda m+z_\perp)\nonumber\\
&&\times\,\mathcal{W}[\lambda m+z_\perp\,;\,\pm\infty m+z_\perp]\mathcal{W}[\pm \infty m+z_\perp\,;\,\pm\infty m+\infty z_\perp]\,|0\rangle\,.\nonumber \\ \label{DeltaTMD}
\end{eqnarray}
The kinematical twist-3 correlators $\Phi^{\rho}_\partial(x)$ and $\Delta^{\rho}_\partial(z)$ are then generated by weighting the correlators $\Phi(x,k_T)$ and $\Delta(z,p_\perp)$ by $k_T$ or $p_\perp$ and integrating over it,
\begin{eqnarray}
\Phi^{q,\rho}_{\partial,ij}(x)& = & \int d^2 k_T\, k_T^\rho\, \Phi^q_{ij}(x,k_T)\,,\label{partPhix}\\
\Delta^{q,\rho}_{\partial,ij}(z) & = & \int d^2p_\perp\,p^\rho_\perp\, \Delta^q_{ij}(z,p_\perp)\,.\label{partDeltaz} 
\end{eqnarray}
One finds these matrix elements can be parameterized as~\cite{Mulders:1995dh,Boer:1997mf,Goeke:2005hb},
\begin{eqnarray}
\Phi_{\partial}^{q,\rho}(x) & \!=\! & \frac{1}{2}\Bigg(M\epsilon^{Pn\rho S}\slash P\, f_{1T}^{\perp(1),\, q}(x) -M(S^{\rho}-P^{\rho}(n\cdot S))\,\slash P\gamma_{5}\, g_{1T}^{(1),\, q}(x)\nonumber \\
 &  & +\frac{1}{2}M^{2}(n\cdot S)\,\left([\slash P,\gamma^{\rho}]\gamma_{5}-P^{\rho}[\slash P,\slash n]\gamma_{5}\right)\, h_{1L}^{\perp(1),\, q}(x) \nonumber\\
 && +\frac{i}{2}M\,\left([\slash P,\gamma^{\rho}]-P^{\rho}[\slash P,\slash n]\right)\, h_{1}^{\perp(1),\, q}(x)\Bigg)\,,\label{eq:ParamPhipart}
\end{eqnarray}
\begin{eqnarray}
\Delta_{\partial}^{q,\rho}(z) & \!=\! & \frac{1}{z}\Bigg(M_{h}\epsilon^{P_{h}m\rho S_{h}}\slash P_{h}\, D_{1T}^{\perp(1),\, q}(z)\nonumber\\
&&-M_{h}(S_{h}^{\rho}-P_{h}^{\rho}(m\cdot S_{h}))\,\slash P_{h}\gamma_{5}\, G_{1T}^{\perp(1),\, q}(z)\nonumber \\
 &  & +\frac{1}{2}M_{h}^{2}(m\cdot S_{h})\,\left([\slash P_{h},\gamma^{\rho}]\gamma_{5}-P_{h}^{\rho}[\slash P_{h},\slash m]\gamma_{5}\right)\, H_{1L}^{\perp(1),\, q}(z)\nonumber\\
 && +\frac{i}{2}M_{h}\,\left([\slash P_{h},\gamma^{\rho}]-P_{h}^{\rho}[\slash P_{h},\slash m]\right)\, H_{1}^{\perp(1),\, q}(z)\Bigg)\,.\label{eq:ParamDeltapart}
\end{eqnarray}
where $f^{(1)}(x)$ and $D^{(1)}(z)$ are defined in Eqs.~(\ref{e:kTmom}), (\ref{e:pTmom}).

We finish with the dynamical functions, which are quark-gluon-quark correlators given by
\begin{eqnarray}
\Phi^{q,\rho}_{F,ij}(x,x_1) & \!=\! & \int_{-\infty}^\infty\frac{d\lambda}{2\pi}\int_{-\infty}^\infty\frac{d\mu}{2\pi}\,\mathrm{e}^{ix_1 \lambda+i(x-x_1)\mu}\nonumber\\
&&\hspace{-1.6cm}\times\,\langle P,S|\,\bar{q}_j(0)\,\mathcal{W}[0\,;\,\mu n]\,ign_{\eta}F^{\eta \rho}(\mu n)\,\mathcal{W}[\mu n\,;\,\lambda n]\,q_i(\lambda n)\,|P,S\rangle\,, \label{PhiFxx'}\\[0.3cm]
\Delta^{q,\rho}_{F,ij}(z,z_1)& \!=\!& \frac{1}{N_c}\sum_{X}\hspace{-0.5cm}\int  \int_{-\infty}^\infty\frac{d\lambda}{2\pi}\int_{-\infty}^\infty\frac{d\mu}{2\pi}\,\mathrm{e}^{i\frac{\lambda}{z_1}+i(\frac{1}{z}-\frac{1}{z_1})\mu} \hspace{-0cm}\langle 0|\,\mathcal{W}[\pm \infty m\,;\,\mu m]\,igm_\eta F^{\eta \rho}(\mu m)  \nonumber\\
&& \times\,\mathcal{W}[\mu m\,;\, \lambda m] q_i(\lambda m)\,|P_hS_h\,;\,X\rangle \hspace{-0cm}\langle P_hS_h\,;\,X|\,\bar{q}_j(0)\,\mathcal{W}[0\,;\, \pm\infty m]\,|0\rangle\,.\nonumber\\ \label{DeltaFzz'}
\end{eqnarray}
These matrix elements can likewise be parameterized in terms of scalar functions, and they read~\cite{Efremov:1981sh,Qiu:1991wg,Eguchi:2006qz,Meissner:2009,Zhou:2009jm,Metz:2012ct,Kanazawa:2013uia,Kanazawa:2015jxa,Kanazawa:2015ajw}
\begin{eqnarray}
\Phi_{F}^{q,\rho}(x,x_1) & = & \frac{M}{2}\Bigg(\epsilon^{Pn\rho S}\,\slash P\, iF_{FT}^{q}(x,x_1)-(S^{\rho}-P^{\rho}(n\cdot S))\,\slash P\gamma_{5}\, G_{FT}^{q}(x,x_1)\nonumber \\
 &  & +\frac{i}{2}\left([\slash P,\gamma^{\rho}]-P^{\rho}[\slash P,\slash n]\right)\, iH_{FU}^{q}(x,x_1)\nonumber \\
 &  & +\frac{M}{2}(n\cdot S)\,\left([\slash P,\gamma^{\rho}]\gamma_{5}-P^{\rho}[\slash P,\slash n]\gamma_{5}\right)\, H_{FL}^{q}(x,x_1)\Bigg)\,,\label{eq:ParamPhiF}
\end{eqnarray}
\begin{eqnarray}
\Delta_{F}^{q,\rho}(z,z_1) & = & \frac{M_{h}}{z}\Bigg(\epsilon^{P_{h}m\rho S_{h}}\slash P_{h}\,i\hat{D}_{FT}^{q}(z,z_1)-(S_{h}^{\rho}-P_{h}^{\rho}(m\cdot S_{h}))\,\slash P_h \gamma_5\,\hat{G}_{FT}^{q}(z,z_1)\nonumber \\
 &  & +\frac{i}{2}([\slash P_{h},\gamma^{\rho}]-P_{h}^{\rho}[\slash P_{h},\slash m])\,i\hat{H}_{FU}^{q}(z,z_1)\nonumber \\
 &  &+\frac{M_h}{2}(m\cdot S_{h})\,([\slash P_{h},\gamma^{\rho}]\gamma_{5}-P_{h}^{\rho}[\slash P_{h},\slash m]\gamma_{5})\,\hat{H}_{FL}^{q}(z,z_1)\Bigg)\,.\label{eq:ParamDeltaF}
\end{eqnarray}
The dynamical PDFs have certain symmetry properties:~$F_{FT}(x,x_1)=F_{FT}(x_1,x)$,
$H_{FU}(x,x_1)=H_{FU}(x_1,x)$, $G_{FT}(x,x_1)=-G_{FT}(x_1,x)$
and $H_{FL}(x,x_1)=-H_{FL}(x_1,x)$. Hence, $G_{FT}(x,x)=H_{FL}(x,x)=0$.
In addition, they have support on $|x\,| \le 1$, $|x_1|\le1$, and $|x-x_1|\le1$.  On the fragmentation
side we have in general $\Delta_{F}^{\rho}(z,z)=0$~\cite{Meissner:2008yf}, $\Delta_{F}^{\rho}(z,0)=0$~\cite{Meissner:2008yf}, and $\frac{\partial}{\partial (1/z_1)}\Delta_{F}^{\rho}(z,z_1)\Big|_{z_1 = z}=0$\cite{Kanazawa:2015ajw}. 
Also, the dynamical FFs are complex and have support on $0\le z\le1$ and $z<z_1<\infty$.

We mention that the dynamical functions in Eqs.~(\ref{eq:ParamPhiF}), (\ref{eq:ParamDeltaF}) are sometimes called F-type functions because they are defined using the gluon field strength tensor $F^{\mu\nu}$.  There are also so-called D-type functions where one makes the replacement $gn_{\eta}F^{\eta \rho} \to D^{\rho}$ (or $gm_{\eta}F^{\eta \rho} \to D^{\rho}$), with $D^\rho = \partial^\rho -igA^\rho$ the covariant derivative.  However, the D-type functions can be written in terms of the F-type (see \ref{a:FD}), so they are not additional independent functions.  We also note that one can define antiquark/charge-conjugated twist-3 PDFs and FFs, and we refer the reader to Ref.~\refcite{Kanazawa:2015ajw} for a nice summary of these functions and (for the PDF case) their relations to the quark ones.

\section{Operator relations between collinear twist-3 functions}

\subsection{Relations between F-type and D-type functions} \label{a:FD}
As we discussed in the last paragraph of \ref{a:def}, the dynamical twist-3 functions can be defined in terms of either F-type correlators (that involve the gluon field strength tensor) or D-type correlators (that involve the covariant derivative).  Here we give the relations between these two types of functions.  For the twist-3 PDFs we have\footnote{Note that $PV$ is the principal value.} (see also Refs.~\refcite{Efremov:1981sh,Qiu:1998ia,Eguchi:2006qz,Koike:2015zya,Koike:2016ura,Zhou:2009jm})
\begin{align}
 F^q_{DT}(x,x_1) &= PV \frac{F^q_{FT}(x,x_1)}{x-x_1} \,,
\label{e:DF_1}
\\[0.3cm]
 G^q_{DT}(x,x_1) &= PV \frac{G^q_{FT}(x,x_1) }{x-x_1} + \delta(x-x_1) \, g_{1T}^{(1),q}(x)\,,
\label{e:DF_2} \\[0.3cm]
H^q_{DL}(x,x_1) &= PV \frac{H^q_{FL}(x,x_1)}{x-x_1} + \delta(x-x_1) \, h_{1L}^{\perp(1),q}(x)\,,
\label{e:DF_4} \\[0.3cm]
 H^q_{DU}(x,x_1) &= PV \frac{H^q_{FU}(x,x_1)}{x-x_1} \,,
\label{e:DF_3}
\end{align}
and for the twist-3 FFs we find (see also Refs.~\refcite{Kang:2010zzb,Metz:2012ct,Kanazawa:2013uia,Kanazawa:2015jxa})
\begin{align}
 \hat{D}^q_{DT}(z,z_1) &= \frac{\hat{D}^q_{FT}(z,z_1)}{\frac{1} {z}-\frac{1} {z_1}} - i\,\delta(1/z - 1/z_1)\,D_{1T}^{\perp(1),q}(z)\,,
\label{e:DF_1FF}
\\[0.3cm]
\hat{G}^q_{DT}(z,z_1) &= \frac{\hat{G}^q_{FT}(z,z_1)}{\frac{1} {z}-\frac{1} {z_1}} + \delta(1/z - 1/z_1)\,G_{1T}^{(1),q}(z)\,,
\label{e:DF_2FF} \\[0.3cm]
\hat{H}^q_{DL}(z,z_1) &= \frac{ \hat{H}^q_{FL}(z,z_1)}{\frac{1} {z}-\frac{1} {z_1}} +\,\delta(1/z - 1/z_1)\,H_{1L}^{\perp(1),q}(z)\,,
\label{e:DF_4FF} \\[0.3cm]
\hat{H}^q_{DU}(z,z_1) &= \frac{\hat{H}^q_{FU}(z,z_1)}{\frac{1} {z}-\frac{1} {z_1}} - i\,\delta(1/z - 1/z_1)\,H_{1}^{\perp(1),q}(z)\,.
\label{e:DF_3FF}
\end{align}

\subsection{Equation of motion relations} \label{a:EOM}
One can use the QCD equation of motion (EOM),\footnote{The quark mass is denoted $m_q$.}~$(i\,\slash D (y) -m_q)q(y)=0$, to connect the intrinsic, kinematical, and dynamical functions.  These are called EOM relations and read (see also Refs.~\citen{Efremov:1981sh,Boer:1997bw,Bacchetta:2006tn,Eguchi:2006qz,Meissner:2009,Zhou:2009jm,Metz:2012ct,Kanazawa:2013uia,Kanazawa:2015jxa})
\begin{align}
x\,g_T^q(x)&=g_{1T}^{(1),q}(x)+\frac{m_q}{M}\,h_1^q(x)-
PV\!\!\int_{-1}^1\!\!dx_1 \,\frac{F_{FT}^q(x,x_1)-G_{FT}^q(x,x_1)}{x-x_1},\label{EoMgT} \\[0.3cm]
x\,h_L^q(x) &= -2h_{1L}^{\perp (1),q}(x)+\frac{m_q}{M}\,g_1^q(x)-2\,PV\int_{-1}^1dx_1 \,\frac{H_{FL}^q(x,x_1)}{x-x_1},\label{EoMhL} \\[0.3cm]
x\,e^q(x)&=\frac{m_q}{M}\,f_1^q(x)-2\,PV\int_{-1}^1dx_1\,\frac{H_{FU}^q(x,x_1)}{x-x_1},\label{EoMe}
\end{align}
for the twist-3 PDFs, and
\begin{align}
\frac{1} {z} \left(iH^q(z)-E^q(z)\right) &= 2\int_z^\infty\frac{dz_1} {z_1^2}\frac{\hat{H}^q_{FU}(z,z_1) } {\frac{1} {z}-\frac{1} {z_1}}- 2iH_1^{\perp(1),q}(z) - \frac{m_q} {M_h}\,D_1^q(z)\,, \label{e:EOM1FF}\\[0.3cm]
\frac{1} {z} \left(iD_T^q(z)-G_T^q(z)\right)& = \int_z^\infty\frac{dz_1} {z_1^2}\frac{\left(\hat{D}^q_{FT}(z,z_1)-\hat{G}_{FT}^q(z,z_1)\right)} {\frac{1} {z}-\frac{1} {z_1}}\, \nonumber\\
&\hspace{0.5cm}- \left(iD_{1T}^{\perp(1),q}(z)+G_{1T}^{(1),q}(z)\right) - \frac{m_q} {M_h}\,H_1^q(z)\,, \label{e:EOMFF2} \\
\frac{1} {z} \left(iE_L^q(z)+H_L^q(z)\right) &= -2\int_z^\infty\frac{dz_1} {z_1^2}\frac{\hat{H}^q_{FL}(z,z_1) } {\frac{1} {z}-\frac{1} {z_1}}- 2H_{1L}^{\perp(1),q}(z) + \frac{m_q} {M_h}\,G_1^q(z)\,, \label{e:EOM3FF}
\end{align}
for the twist-3 FFs.

\subsection{Lorentz invariance relations} \label{a:LIR}
Another set of formulae between the intrinsic, kinematical, and dynamical functions are derived from identities among non-local
operators where constraints from Lorentz invariance are taken into account.  These are known as Lorentz invariance relations (LIRs).  The LIRs for twist-3 PDFs read (see also Refs.~\citen{Eguchi:2006qz,Kodaira:1998jn,Belitsky:1997ay,Accardi:2009au})
\begin{align}
 g_T^q(x) &= g_1^q(x) + \frac{d}{dx} g_{1T}^{(1),q}(x) - 2 \, PV
  \int_{-1}^1 \!dx_1 \,\frac{G_{FT}^q(x,x_1)}{(x-x_1)^2}\,, 
  \label{LIRgT}\\[0.3cm]
  h_L^q(x) &= h_1^q(x) - \frac{d}{dx} h_{1L}^{\perp (1),q}(x)
  + 2 \, PV \int_{-1}^1 \!dx_1 \,\frac{H_{FL}^q(x,x_1)}{(x-x_1)^2}\,,
 \label{LIRhL}
\end{align}
and for twist-3 FFs are (see Ref.~\refcite{Kanazawa:2015ajw} where these expressions were derived for the first time)
\begin{align}
\frac{D_T^q(z)}{z} &= - \left( 1- z \frac{d}{dz} \right) D_{1T}^{\perp
(1),q}(z) - \frac{2}{z} \int_z^\infty \frac{dz_1}{z_1^2}
 \frac{\hat{D}_{FT}^{q,\Im}(z,z_1)}{(1/z-1/z_1)^2}\,, \label{LIRDT} \\[0.3cm]
 \frac{G_T^q(z)}{z} &= \frac{G_1^q(z)}{z} 
  + \left( 1 - z\frac{d}{dz} \right) G_{1T}^{(1),q} (z)  - \frac{2}{z}
  \int_z^\infty \frac{dz_1}{z_1^2}
  \frac{\hat{G}_{FT}^{q,\Re}(z,z_1)}{(1/z-1/z_1)^2}\,, \label{LIRGT}\\[0.3cm]
 \frac{H_L^q(z)}{z} &= \frac{H_1^q(z)}{z} - \left( 1 - z \frac{d}{dz} \right)
 H_{1L}^{\perp(1),q}(z)  + \frac{2}{z} \int_z^\infty
 \frac{dz_1}{z_1^2}
 \frac{\hat{H}_{FL}^{q,\Re}(z,z_1)}{(1/z-1/z_1)^2}\,, 
\label{LIRHL} \\[0.3cm]
  \frac{H^q(z)}{z} &= - \left( 1 - z \frac{d}{dz} \right) H_{1}^{\perp
  (1),q}(z)  - \frac{2}{z} \int_z^\infty \frac{dz_1}{z_1^2}
  \frac{\hat{H}_{FU}^{q,\Im}(z,z_1)}{(1/z-1/z_1)^2}\,. \label{LIRH}
\end{align}

\subsection{Expressions for intrinsic and kinematical twist-3 functions in terms of dynamical twist-3 functions} \label{a:dyn}
In \ref{a:EOM} and \ref{a:LIR} we have seen how intrinsic, kinematical, and dynamical functions are related to each other.  From those formulae, one can solve for the intrinsic and kinematical functions in terms of the dynamical ones.  We find\footnote{Note that $\epsilon(x)=2\theta(x)-1$, and PDFs at $x<0$ represent antiquark distributions.  
Also, it is to be understood that $x$ falls within the range of integration $(x,\epsilon(x))$, i.e.,  $\int_x^{\epsilon(x)} d{x_1}f(x_1)\delta(x_1-x) = f(x)$.  Similarly, $\int_z^{1} d{z_1}D(z_1)\delta(1/z_1-1/z) = z^2D(z)$.}(see Ref.~\refcite{Kanazawa:2015ajw})
\begin{align}
g_T^q(x)&=\int_x^{\epsilon(x)}dx'\,\frac{g_1^q(x')}{x'}
+{m_q\over M}\left( {1\over x}h_1^q(x) +\int^x_{\epsilon(x)}dx' \,{h_1^q(x')\over {x'}^2}\right)\nonumber\\
&\hspace{0cm}+ \int_x^{\epsilon(x)}\frac{dx_1}{x_1^2}\,PV\!\int_{-1}^1dx_2\,\Bigg[\frac{1-x_1\,\delta(x_1-x)}{x_1-x_2}F_{FT}^q(x_1,x_2)\nonumber\\
&-\frac{3x_1-x_2-x_1(x_1-x_2)\,\delta(x_1-x)\,}{(x_1-x_2)^2}G_{FT}^q(x_1,x_2)\Bigg]\,,
\label{gT} \\[0.3cm]
g_{1T}^{(1),q}(x)& = x\int_x^{\epsilon(x)}dx'\,\frac{g_1^q(x')}{x'} +{m_q\over M} x \int^x_{\epsilon(x)}dx'\,{h_1^q(x')\over {x'}^2}\nonumber\\
&\hspace{0cm}+ x\int_x^{\epsilon(x)}\frac{dx_1}{x_1^2}\,PV\!\int_{-1}^1dx_2\,\Bigg[\frac{F_{FT}^q(x_1,x_2)}{x_1-x_2}-\frac{(3x_1-x_2)\,G_{FT}^q(x_1,x_2)}{(x_1-x_2)^2}\Bigg]\,, \label{g1T}\\[0.5cm]
h_L^q(x)&=2x\int_x^{\epsilon(x)}dx_1\,{h_1^q(x_1)\over x_1^2}+{m_q\over M}\left( {g_1^q(x)\over x} -2x \int_x^{\epsilon(x)}dx_1\,{g_1^q(x_1)\over x_1^3}\right)\nonumber\\
&\hspace{-0.85cm}+4x \int_x^{\epsilon(x)}{dx_1\over x_1^3}PV\!\int_{-1}^1dx_2\,{(x_1/2)(x_2-x_1)\delta(x_1-x)+2x_1-x_2 \over (x_1-x_2)^2}H_{FL}^q(x_1,x_2)\,.\nonumber\\ 
\label{hLWW}\\[0.3cm]
h_{1L}^{\perp (1),q}(x)&=x^2\int_x^{\epsilon(x)}dx_1\,{h_1^q(x_1)\over x_1^2}-{m_q\over M}x^2 \int_x^{\epsilon(x)}dx_1\,{g_1^q(x_1)\over x_1^3}\nonumber\\
&\hspace{-0.85cm}+2x^2 \int_x^{\epsilon(x)}{dx_1\over x_1^3}\,PV\!\int_{-1}^1dx_2\,{2x_1-x_2 \over (x_1-x_2)^2}H_{FL}^q(x_1,x_2)\,.  
\label{h1L(1)WW}\\[0.1cm] \nonumber
\end{align}
for the twist-3 PDFs, and

\begin{align}
D_{T}^{q}(z)&=-z\int_z^1\frac{dz_1}{z_1}\int_{z_1}^\infty \frac{dz_2}{z_2^2}\,\times\nonumber\\
&\hspace{-0.5cm}\left[\frac{\left(1+\frac{1}{z_1}\delta\left(\frac{1}{z_1}-\frac{1}{z}\right)\right)\,\hat{G}_{FT}^{q,\Im}(z_1,z_2)}{\frac{1}{z_1}-\frac{1}{z_2}}\right.\nonumber\\
&\left.\hspace{-0.5cm}-\frac{\left(\frac{3}{z_1}-\frac{1}{z_2}+\frac{1}{z_1}\left(\frac{1}{z_1}-\frac{1}{z_2}\right)\,\delta\left(\frac{1}{z_1}
-\frac{1}{z}\right)\right)\,\hat{D}_{FT}^{q,\Im}(z_1,z_2)}{\left(\frac{1}{z_1}-\frac{1}{z_2}\right)^2}\right]\,, \label{DT} \\[0.3cm]
D_{1T}^{\perp (1),q}(z)&=\int_z^1\frac{dz_1}{z_1}\int_{z_1}^\infty \frac{dz_2}{z_2^2}\,
\Bigg[\frac{\hat{G}_{FT}^{q,\Im}(z_1,z_2)}{\frac{1}{z_1}-\frac{1}{z_2}}-\,\frac{\left(\frac{3}{z_1}-\frac{1}{z_2}\right)\,\hat{D}_{FT}^{q,\Im}(z_1,z_2)}{\left(\frac{1}{z_1}-\frac{1}{z_2}\right)^2}\Bigg]\,,\label{D1Tperp1} \\[0.5cm]
G_{T}^{q}(z)&=-z\int_z^1dz'\,\frac{G_1^q(z')}{z'^2}-z\int_z^1\frac{dz_1}{z_1}\int_{z_1}^\infty \frac{dz_2}{z_2^2}\,\nonumber\\
&\hspace{-0.5cm}\times\,\Bigg[\frac{\left(1+\frac{1}{z_1}\delta\left(\frac{1}{z_1}-\frac{1}{z}\right)\right)\,
\hat{D}_{FT}^{q,\Re}(z_1,z_2)}{\frac{1}{z_1}-\frac{1}{z_2}}\nonumber\\
&\hspace{-0.5cm}-\frac{\left(\frac{3}{z_1}-\frac{1}{z_2}+\frac{1}{z_1}\left(\frac{1}{z_1}
-\frac{1}{z_2}\right)\,\delta\left(\frac{1}{z_1}-\frac{1}{z}\right)\right)\,
\hat{G}_{FT}^{q,\Re}(z_1,z_2)}{\left(\frac{1}{z_1}-\frac{1}{z_2}\right)^2}\Bigg]\nonumber\\
&+{m_q\over M_h}\left(z\,H^q_1(z) + z\int_z^1dz'\,\frac{H^q_1(z')} {z'}\right)\, ,
\label{GT}\\[0.3cm]
G_{1T}^{(1),q}(z)&=-\int_z^1dz'\,\frac{G_1^q(z')}{z'^2}-\int_z^1\frac{dz_1}{z_1}\int_{z_1}^\infty \frac{dz_2}{z_2^2} \nonumber\\
&\hspace{-1cm}\times\,\Bigg[\frac{\hat{D}_{FT}^{q\,\Re}(z_1,z_2)}{\frac{1}{z_1}-\frac{1}{z_2}}-\frac{\left(\frac{3}{z_1}-\frac{1}{z_2}\right)\,\hat{G}_{FT}^{q,\Re}(z_1,z_2)}{\left(\frac{1}{z_1}-\frac{1}{z_2}\right)^2}\Bigg]+{m_q\over M_h}\int_z^1dz'\,\frac{H_1^q(z')} {z'}\, ,
\label{G1T1}\\[0.5cm]
H_L^q(z)&=-2\int_z^1 dz'\,{H_1^q(z')\over z'} +{m_q\over M_h} \left(z\,G_1^q(z)+2\int_z^1dz'\,G_1^q(z')\right)\nonumber\\
& -\,4\int_z^1dz_1\int_{z_1}^\infty {dz_2\over z_2^2}\,\hat{H}_{FL}^{q,\Re}(z_1,z_2) \nonumber\\
&\hspace{0.5cm}\times\,{(1/(2z))(1/z_1-1/z_2)\delta(1/z_1-1/z)+2/z_1- 1/z_2 \over (1/z_1-1/z_2)^2 } \,, \label{HLWW} \\[0.3cm]
H_{1L}^{\perp (1),q}(z)&=-{1\over z}\int_z^1dz'\,{H_1^q(z')\over z'}+{m_q\over M_h}{1\over z} \int_z^1dz'\,G_1^q(z')\nonumber\\
& -{2\over z}\int_z^1dz_1\int_{z_1}^\infty \, {dz_2\over z_2^2} {2/z_1 - 1/z_2 \over (1/z_1-1/z_2)^2}
\hat{H}_{FL}^{q,\Re}(z_1,z_2)\,,
\label{H1LWW}\\[0.5cm]
H^{q}(z)& =  \int_z^1dz_1\int_{z_1}^\infty \frac{dz_2}{z_2^2}\nonumber\\
&\times\, 2\,\Bigg[\frac{\left(\,2(\frac{2}{z_1}-\frac{1}{z_2})+\frac{1}{z_1}\left(\frac{1}{z_1}-\frac{1}{z_2}\right)\,
\delta\left(\frac{1}{z_1}-\frac{1}{z}\right)\right)\hat{H}_{FU}^{q,\Im}(z_1,z_2)}{\left(\frac{1}{z_1}-\frac{1}{z_2}\right)^2}\Bigg]\,, \label{H}\\[0.3cm]
H_1^{\perp (1),q}(z)&=-\frac{2}{z}\int_z^1dz_1\int_{z_1}^\infty \frac{dz_2}{z_2^2}\,\frac{\left(\frac{2}{z_1}-\frac{1}{z_2}\right)\,\hat{H}_{FU}^{q,\Im}(z_1,z_2)}{\left(\frac{1}{z_1}-\frac{1}{z_2}\right)^2}\,,
&\label{H1perp1} 
\end{align}
for the twist-3 FFs.  These expressions, along with Eqs.~(\ref{e:QS_Siv}), (\ref{e:HFU_BM}), (\ref{EoMe}), (\ref{e:EOM1FF}) (\ref{e:EOM3FF}), show that multi-parton correlations are the fundamental objects accessed through transverse spin observables in hadronic processes.

\end{document}